\documentclass[a4paper,reprint,aps,unsortedaddress]{revtex4}

\usepackage[utf8]{inputenc}
\usepackage[english]{babel}

\usepackage[hidelinks]{hyperref}

\usepackage{graphicx}
\usepackage{subcaption}
\usepackage{parskip}
\usepackage{appendix}
\usepackage{float}
\usepackage{booktabs}
\usepackage{makecell}
\usepackage{multirow}

\usepackage[figurename=Fig.]{caption}
\usepackage[tablename=Tab.]{caption}
\usepackage[font=small]{caption}

\usepackage{amsmath}
\usepackage{amssymb}
\usepackage{amsfonts}
\usepackage{mathtools}
\usepackage{upgreek}
\usepackage{xcolor}
\usepackage[normalem]{ulem}
\usepackage{cancel}

 \newcommand{\mc}[1]{{\textcolor{violet}{{#1}}}}

\newcommand{\ee}{\end{equation}}
\newcommand{\eea}{\end{eqnarray}}
\newcommand{\be}{\begin{equation}}
\newcommand{\bea}{\begin{eqnarray}}

\begin{document}

\title{Circular orbits and particle collisions close to\\
charged black holes surrounded by scalar clouds}

\date{\today}

\author{Maria Chivers}
 \email{maria.chivers.20@ucl.ac.uk}
\affiliation{Department of Mathematics, University College London, Gower Street, London, WC1E 6BT, UK}

\author{Betti Hartmann}
 \email{b.hartmann@ucl.ac.uk}
\affiliation{Department of Mathematics, University College London, Gower Street, London, WC1E 6BT, UK}

\author{Katherine Horton}
 \email{katherine.horton.22@ucl.ac.uk}
\affiliation{Department of Mathematics, University College London, Gower Street, London, WC1E 6BT, UK}

\author{Yves Brihaye}
 \email{yves.brihaye@umons.ac.uk}
\affiliation{Physique de l'Univers, Universit\'e de
Mons, 7000 Mons, Belgium}

\begin{abstract}
We study the motion of massive (un)charged test particles in space-times of electric and dyonic black holes which carry scalar hair. We determine the stable and unstable circular orbits and discuss the collision of massive test particles. In particular, we aim at demonstrating how the presence of the scalar hair of the black hole changes the circular orbits and particle collisions, respectively, as compared to the Reissner-Nordstr\"om (RN) space-time. We find that in the presence of scalar hair, up to four circular orbits (two unstable and two stable) as well as static orbits with $L=0$ can exist. Particle collisions can generate infinite center-of-mass energy when at least one of the particles is charged, very similar to the RN case. We find, however, that the value of the charge at which this divergence happens depends on the value of the scalar field on the horizon. 

\end{abstract}

\maketitle

\section{Introduction}
To endow black holes with scalar hair is not straightforward - various no-go theorems demonstrate this \cite{nohair,noscalarhair}. When considering 
asymptotically flat, stationary rotating black holes of the Kerr-type, this is possible when considering a complex massive scalar field \cite{scalar_Kerr}.
For static, spherically symmetric black holes, it turns out that the complex scalar field needs to be charged and possess scalar self-interactions beyond the quartic order \cite{Hong:2019mcj,Herdeiro:2020xmb,Brihaye:2020vce,Hong:2020miv}. These solutions typically feature a spherically symmetric electric field analogous to that of the Reissner–Nordstr\"om black hole. More recently, a new class of spherically symmetric hairy black holes has been introduced \cite{Herdeiro:2024yqa,Brihaye:2025nfx}, characterized by the presence of both electric and magnetic charges. The construction relies on the inclusion of a multiplet of complex scalar fields with a global $SU(n+1)$ symmetry, together with a carefully chosen angular dependence that preserves cohomogeneity-one field equations. The coupling $g$ between the scalar fields and a magnetic monopole with charge $Q_m$ requires the Dirac quantization condition
$g Q_m = \pm \frac{N}{2}$ with $N$ an integer. Although the magnetic monopole field is singular at the origin, this singularity is hidden behind the event horizon.
Given that these solutions exist in gravity-scalar-gauge field model without modifications of gravity and/or the electrodynamical interaction, the question remains if the scalar hair on black holes can actually be detected. One possibility of detection is the observation of test particle motion in the space-times of these objects. In the following we will concentrate on massive uncharged and charged test particles, the former moving on geodesics. 

The innermost stable circular orbit (ISCOs) is defined as the smallest marginally stable circular orbit on which a massive test particle can orbit a compact object. 
Circular orbits lying inside the ISCO are unstable and hence any small perturbation of the particle motion would lead to the particle plunging into the black hole. When modeling accretion disks around black holes, the ISCO is believed to define the inner edge of the disk \cite{isco_accretion}. 
The radius of the ISCO of the Schwarschild black hole is at $r_{\rm ISCO}=6M$, where $M$ is the ADM mass of the solution. For Reissner-Norstr\"om black holes,
circular geodesic motion was studied in \cite{Pugliese:2010ps}, which was extended to the case of charged particles in \cite{Pugliese:2011py, Schroven:2020ltb}.
The discussion of the general motion of electrically and magnetically charged particles in the Reissner-Nordstr\"om space-time (including a magnetic charge for the black hole) was done in \cite{Grunau:2011gd}. ISCOs have also been studied in a number of black hole space-times that contain scalar fields. Einstein-dilaton-Gauss-Bonnet theory has black hole solutions
and massive test particle motion was discussed e.g. in \cite{Pani:2009wy,Maselli:2015tta,Blazquez-Salcedo:2016yka}. Rotating black holes can carry complex scalar hair when the so-called ``synchronisation condition'' is fulfilled and the ISCOs in the space-time of these black holes have been discussed \cite{Herdeiro:2014goa}. This was also done for a black hole solution
in a generalized scalar-tensor gravity theory \cite{Sotiriou:2014pfa}, in a Einstein-Maxwell-scalar field model with non-trivial coupling between the electromagnetic and scalar field \cite{Turimov:2020fme} and for Kerr-like black holes with minimally coupled scalar field \cite{Bogush:2022hop}.

The study of ISCOs is, in fact, related to that of particle collisions \cite{Harada:2014vka}. The collision of massive test particles close to the event horizon of a black hole 
can lead to large center-of-mass energies. 
Ba\~nados, Silk and West (BSW) have explored the possibility of particles achieving infinite center-of-mass energy ($E_{\text{\tiny{C.M.}}}$) due to extreme gravitational environment near a rotating black hole, specifically the Kerr black hole \cite{Banados:2009}.
This infinite energy would, however, not be detectable for an external observer \cite{Jacobson:2010} and hence would not be a possible explanation for Ultra-High-Energy Cosmic rays. Nevertheless, understanding particle interactions in the vicinity of black hole horizons remains important as it tests extreme gravitational fields and their effects on the surrounding matter. As such it has e.g. been suggested that the high energies in particle collisions could alter cross sections for the annihilation of dark matter \cite{Baushev:2009}. 
Rotating black holes are not the only space-time backgrounds in which infinite center-of-mass energy can be achieved. 
As shown in \cite{Zaslavskii:2010}, this can also be achieved without rotation, but in a Reissner-Nordstr\"om (RN) space-time with charged particles. 
It was also pointed out that for non-extremal RN black holes, the energy in the center-of-mass system is finite but can be made as large as possible. In the extremal case, and with charge of the particles $q=Er_h/Q$, where $E$ is the particle's energy and $Q$ the charge of the RN black hole, the center-of-mass energy diverges at the horizon $r=r_h$. In \cite{Zaslavskii:2016}, collisions of spinning particles in the space-time of a Schwarzschild black hole were shown to lead to the center-of-mass energy becoming infinite provided that the particles carry a spin and that there is a restriction on the energy-to-mass ratio. Further studies have been conducted on charged, rotating black holes, revealing that the $E_{\text{\tiny{C.M.}}}$ of collisions between two uncharged particles falling from rest at infinity is influenced not just by the spin of the black hole, but also by its charge \cite{Wei:2010}. In essentially all studies on particle collisions, particles that are at rest at infinity have been considered. In \cite{Hackmann:2020ogy}, on the other hand, it was assumed that one particle involved in the collision resides close to the black hole event horizon in a spherically symmetric, static space-time. 
Particle collisions in black hole space-times with scalar hair have also been discussed \cite{Sultana:2015dda, Sultana:2015avz, Zaslavskii:2016stw}.

In this paper, we are interested in (a) the circular orbits of (un)charged test particles in the space-time of electrically or dyonically charged black holes that carry scalar hair and (b) the collision of (un)charged  test particles close to the event horizon of these black holes. In particular, we will be interested in how the values of the ISCO radius and the center-of-mass energy 
compare to the RN limit.

\section{The field theoretical model and its solutions}
\label{sec:solutions}
The field theoretical model for the electrically and dyonically, respectively, charged black holes with scalar hair has been studied in \cite{Herdeiro:2024yqa} for a sextic potential and in \cite{Brihaye:2025nfx} with a bounded exponential potential. In the following, we are interested in the motion of test particles in the space-time of the solutions found in \cite{Brihaye:2025nfx}. In order to be able to understand the qualitative and quantitative features of the test particle motion, it is crucial to understand the features of the solutions themselves. Hence, we briefly review the field theoretical model here. The action is~:
\begin{equation}
\label{eq:action}
{\cal S} = \int {\rm d}^4 x \sqrt{-g} \left(\frac{{\cal R}}{4\alpha} + {\cal L}_{m}\right) \ ,
\end{equation}
where $\alpha=4\pi G$ and the matter Lagrangian density given by~:
\begin{equation}
\label{eq:matter_lagrangian}
{\cal L}_{m}= \sum\limits_{k=1}^{n+1} \left[-D_{\mu} \Phi_k (D^{\mu}\Phi_k)^{*} -U(\Phi_k\Phi_k^*)\right] -\frac{1}{4} F_{\mu\nu} F^{\mu\nu} \ \ .
\end{equation}
Here $D_{\mu} = \partial_{\mu} - i g A_{\mu}$ is the covariant derivative with gauge coupling $g$, $F_{\mu\nu}=\partial_{\mu}A_{\nu} - \partial_{\nu}A_{\mu}$ is the U(1) field strength tensor and the scalar field potential reads~:
\be
\label{eq:susy_pot}
             U(\Phi_k\Phi_k^*) = m^2 \eta^2 \Bigr(1 - \exp(- \Phi_k\Phi_k^* / \eta^2) \Bigl) \ .
\ee
$\eta$ is an energy scale. The spherically symmetric solutions have metric tensor given by the following line element~:
\be
\label{eq:metric}
     {\rm d}s^2 = - N(r) \sigma^2(r) {\rm d}t^2 + \frac{1}{N(r)} {\rm d}r^2 + r^2 {\rm d}\theta^2 + r^2 \sin^2\theta {\rm d}\varphi^2  \ \ , \ \  N(r)=1-\frac{2\mu(r)}{r} \ ,
\ee
with $\mu(r)$ the mass function and \cite{Herdeiro:2024yqa}
\be
\label{electromagnetic}
               A_{\mu} {\rm d}x^{\mu} = V(r) {\rm d}t + Q_m \cos (\theta) {\rm d} \varphi
\ee
for the U(1) gauge field. The parameter $Q_m$ denotes the magnetic charge of the solution. For the scalar field, we distinguish the cases $k=0$ and $k=1$, where $k \in {\mathbb{N}}$ appears in the Dirac quantisation condition ($c=\hbar=1$)~:
\be
                  g Q_m = \pm \frac{k}{2} \ \ , \ \ k \in \mathbb{N} \ .
\ee
The electric and magnetic fields of the solutions is given by~:
\begin{equation}
E_r=-F_{r0}=-\frac{{\rm d} V}{{\rm d}r} \ \ , \ \   B_r=\frac{F^{\theta\varphi}}{\sqrt{-g}}  = -\frac{Q_m}{r^2} \ ,
\end{equation}
where the latter describes the field of a magnetic monopole located at the origin. 
In the following, the choice of Ansatz for the scalar field depends on whether we choose $Q_m=0$ or $Q_m\neq 0$. For $k=0$, i.e. scalar fields charged only electrically, we choose~:
\be
 \Phi = \phi(r) e^{-i\omega t} 
 \ee
 and for $k=1$, i.e. electrically and magnetically charged scalar fields, we choose \cite{Herdeiro:2024yqa}~:
 \be
\Phi_1 = \phi(r)  \sin\left(\frac{\theta}{2}\right) e^{i( \varphi/2 -\omega t)}\ \  \ , \ \ \  
		\Phi_2 = \phi(r)  \cos\left(\frac{\theta}{2}\right) e^{-i( \varphi/2 +\omega t)} \ .
\ee
The equations of motion that result from the variation of (\ref{eq:action}) are invariant under the following rescaling of the fields and coupling constants
\be
  \phi \rightarrow \frac{\phi}{\eta} \ \ , \ \ r \rightarrow m r \ \ , \ \ V \rightarrow \frac{V}{\eta} \ \ , \ \ 
	 Q_m \rightarrow \frac{m}{\eta} Q_m \ \ , \ \ g \rightarrow \frac{\eta}{m} g \ \ ,
\ee
which, in particular, leaves the Dirac quantization invariant. We will use these
rescalings to set the scalar field mass $m=1$, $\eta=1$ without loss of generality.
As pointed out in \cite{Herdeiro:2020xmb}, in order to have non-trivial scalar fields on a spherically symmetric, charged black hole, the following synchronisation condition needs to be fulfilled~:
\begin{equation}
\label{eq:synchro}
\omega=g V(r_h)\ .
\end{equation}
In the following, we will make the gauge choice $\omega=0$, i.e. $V(r_h)=0$.
With this choice, the field equations read~:
 
\begin{eqnarray}
\mu' &=& \alpha \left[r^2 N \phi'^2 + \frac{r^2 g^2 V^2 \phi^2}{\sigma^2 N} + \frac{k}{2}\phi^2 + \frac{r^2 V'^2}{2\sigma^2} + \frac{k^2}{8g^2 r^2} + r^2 \left(1-\exp(-\phi^2)\right)\right]
\label{eq:m}  \ , \\
\frac{\sigma'}{\sigma} &=& 2\alpha r\left[\phi'^2 + \frac{g^2 V^2 \phi^2}{\sigma^2 N^2}\right]
\label{eq:sigma}  \ , \\
\left(\frac{r^2 V'}{\sigma}\right)' &=& 2 \frac{r^2 g^2 V \phi^2}{\sigma N}\label{eq:V}   \ , \\
\left(\sigma r^2 N \phi'\right)' &=& -\frac{r^2 g^2 V^2 \phi}{\sigma N} + \frac{k}{2} \sigma \phi + r^2 \sigma \phi \exp(-\phi^2) \  .
\label{eq:phi}  
\end{eqnarray}

\begin{figure}[h!]
\begin{center}
{\includegraphics[width=8cm]{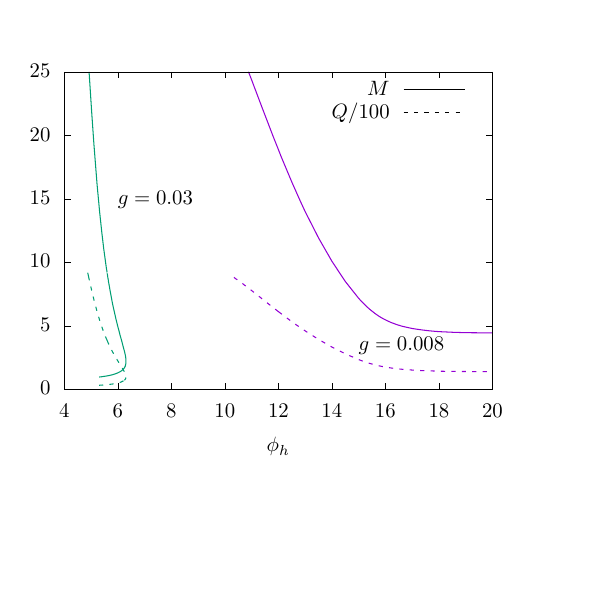}}
{\includegraphics[width=8cm]{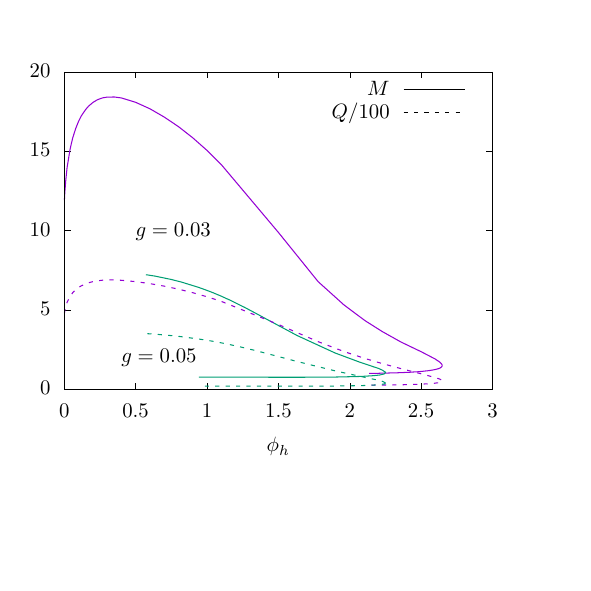}}
\vspace{-1.5cm}
\caption{{\it Left}: We show the dependence of the ADM mass $M$ and the electric charge $Q$ for electrically charged black holes with scalar hair on $\phi(r_h)=\phi_h$ for $k=0$ and two different values of $g$. {\it Right}: Same as left, but for $k=1$. For all solutions we have chosen $\alpha=0.001$ and $r_h=1$.
\label{fig:M_Q_vs_phirh}
}
\end{center}
\end{figure}

These equations have to be solved numerically subject to appropriate boundary conditions. These are~:
\begin{eqnarray}
& & N(r_h)=0 \ \ , \ \  (N'\phi')\vert_{r=r_h}=\frac{k}{2}\sigma(r_h)\phi(r_h) + r^2 \sigma(r_h) \phi(r_h) \exp(-\phi(r_h)^2) \ \ , \nonumber \\
& & V(r_h)=0 \ \ , \ \ \phi(r\rightarrow \infty)\rightarrow 0  \ \ , \ \ 
\sigma(r\rightarrow \infty)\rightarrow 1 \ \ , \ \ V(r\rightarrow \infty) \rightarrow V_{\infty} \  .
\end{eqnarray}
In \cite{Brihaye:2025nfx}, the solutions describing black holes with scalar hair were discussed and we refer the reader for more details to this paper. In order to be able to make the connection between the test particle motion and the field theoretical solutions, we give the ADM mass $M$ and the electric charge $Q$ of the black hole solutions carrying scalar hair in function of the value of the scalar field on the horizon $\phi(r_h)=\phi_h$ in Fig. \ref{fig:M_Q_vs_phirh}. On the left (right), we show the results for electrically (respectively, electrically and magnetically) charged black holes for two different values of the gauge coupling $g$ and for the choice $r_h=1$ and $\alpha=0.001$. As was pointed out already in \cite{Brihaye:2025nfx} a number of branches of solutions exists. In the following it turns out that the solutions for intermediate values of the gauge coupling are the most interesting, hence our choice of values of $g$ for the two cases.

There are a number of limiting cases for the above set of equations. In the following, we will be interested to compare the fully backreacted charged black hole carrying scalar hair with the Reissner-Nordstr\"om solution. This is an explicit solution to the equations (\ref{eq:m}) - (\ref{eq:phi}) for $\phi\equiv 0$ and reads~:
\begin{equation}
\label{eq:RN}
V(r)=V_{\infty} - \frac{Q}{r} \ \ , \ \ \sigma(r)\equiv 1 \ \ , \ \
N(r)=1-\frac{2M}{r} + \frac{\alpha(Q^2 + Q_m^2)}{r^2} \ ,
\end{equation}
where $Q$ is the electric charge, $Q_m=k/(2g)$ is the magnetic charge and $M$ the ADM mass of the solution. For the RN limit the synchronisation condition implies that $V_{\infty}=Q/r_h$. The horizons of the RN solution are~:
\begin{equation}
r_{\pm}=M \pm \sqrt{M^2-\alpha(Q^2 + Q_m^2)} \ ,
\end{equation}
where $r_+=r_h$ is the event horizon and $r_-$ is the Cauchy horizon. For $M=\sqrt{\alpha(Q^2 +Q_m^2)}$ the solution becomes extremal such that $M=r_h$.

\section{Motion of massive test particles}
In the following we will discuss the motion of massive test particles that might be
electrically charged. The Hamiltonian-Jacobi equation describing this motion reads \cite{Grunau:2011gd}~:
\begin{equation}
\label{eq:S}
\frac{\partial S}{\partial\tau}=\frac{1}{2}g^{\mu\nu}\left(\frac{\partial S}{\partial x^{\mu}}-qA_{\mu}\right)\left(\frac{\partial S}{\partial x^{\nu}}-qA_{\nu}\right) \ ,
\end{equation}
where $q$ is the electric charge of the test particle. This can be solved by~:
\begin{equation}
S=-\frac{1}{2}\tau-Et+L_z\varphi+S_r(r)+S_{\theta}(\theta) \ \ ,
\end{equation}
where $E$ is the energy and $L_z$ is the angular momentum in $z$-direction of the charged particle. $S_r$ and $S_{\theta}$ are two functions that depend only on $r$ and $\theta$, respectively. Hence, the components of the equation of motion read~:
\begin{eqnarray}
\label{eq:geodesic_components}
\dot{t}^2=\left(\frac{E+qV}{N\sigma^2}\right)^2 \ \ \ &,& \ \ \ \dot{\varphi}^2=\left(\frac{L_z-qQ_m\cos\theta}{r^2\sin^2\theta}\right)^2 \nonumber \\
\dot{r}^2=\frac{(E+qV)^2}{\sigma^2}-N\bigg(1+\frac{C}{r^2}\bigg) \ \ &,& \ \ \dot{\theta}^2=\frac{1}{r^4}\left(C-\frac{(L_z-qQ_m\cos\theta)^2}{\sin^2\theta}\right) \ ,
\end{eqnarray}
where $C$ is a separation constant defined as $C=(r^2 \dot{\theta})^2+(L_z-qQ_m\cos\theta)^2/\sin^2\theta$. The meaning of this constant becomes clear, when remembering that the mechanical angular momentum $L_{\text{mech}}^2=(L^{\text{mech}}_x)^2+ (L^{\text{mech}}_y)^2 +(L^{\text{mech}}_z)^2=r^4\dot{\theta}^2+r^4\sin^2\theta\dot{\varphi}^2$. From the equations of motion for \(\dot{\varphi}\), we have constant $L_z$ which is the canonical conserved momentum appearing in the action and is defined as \(L_z=p_{\varphi}=g_{\varphi\varphi}\dot{\varphi}+qA_{\varphi}=r^2\sin^2\theta\dot{\varphi}+qQ_m\cos\theta\), whereas \(L^{\text{mech}}_z=L_z-qQ_m\cos\theta\). Since the canonical \(L_z\) is constant, the only way $L^{\text{mech}}_z$ will change is if the particle moves in $\theta$ direction. Hence, if the particle's motion is not confined to the equatorial plane, \(d(L^{\text{mech}}_z)/dt\neq0\). This change comes from the torque exerted on the charged test particle by the magnetic field of the black hole. However, both of the quantities are conserved if the motion is restricted to the equatorial plane and so $L^{\text{mech}}_z=L_z= L$. 


 Using the normalization $\dot{x}_{\mu}\dot{x}^{\mu}=-1$ we find from (\ref{eq:geodesic_components}) that
\begin{equation}
E= \sigma\sqrt{ \dot{r}^2 + N  r^2\dot{\theta}^2 + \frac{N(L    -q Q_m \cos\theta)^2}{r^2\sin^2\theta} + N} - q V \ ,
\end{equation}
such that the energy for a particle at rest at infinity is \(E=1-qV_{\infty}\), which due to our choice of gauge is non-equal to unity for $q\neq 0$. We will choose $\theta=\pi/2$ in the following, which can always be done when $q=0$ and is our choice for $q\neq 0$. The equation for the radial motion then reads~:
\begin{equation}\label{eq:rdot}
\dot{r}^2+ V_{\rm eff}(r)=0 \ \ , \ \ V_{\rm eff}=-\frac{(E+qV)^2}{\sigma^2}+N\bigg(1+\frac{L^2}{r^2}\bigg)  \ .
\end{equation}
In the following we will be interested in circular orbits as well as the collision of massive particles in the space-time determined numerically by solving (\ref{eq:m}) - (\ref{eq:phi}) subject to appropriate boundary conditions using an adaptive Newton-Raphson method \cite{colsys}.

Circular orbits occur at stationary points of the effective potential $V_\text{eff}$, i.e. for $V_{\rm eff}(r)=0$ and  $V'_{\text{eff}}(r)=0$, where the prime denotes the derivative with respect to $r$. Using the effective potential defined in (\ref{eq:rdot}), the conditions
$V_{\rm eff}=0$ and $V'_{\rm eff}=0$ lead to the angular momentum of the circular orbit to be~:
\begin{eqnarray}
\label{eq:lz2}
    L^2 &=& -\frac{1}{\Gamma^2} \left[\left(2((2 r\sigma' - \sigma)N' \sigma - q^2 r V'^2)N +  4(r\sigma' - \sigma) N^2\sigma' + N'^2 r\sigma^2\right) r^3 \right. \nonumber \\
    &\pm& \left. 2\sqrt{ - 4 N r \sigma' \sigma + 4 N \sigma^2 - 2 N' r \sigma^2 + q^2 r^2 V'^2}  N q r^3 V'\right] \ \ \ , \ \  \Gamma=2r N\sigma' - 2N\sigma + N'r \sigma  \ .
\end{eqnarray}
The circular orbits typically arise in minimum-maximum pairs, with the stable circular orbit corresponding to the minimum of the effective potential ($V''_{\rm eff} > 0$) and having greater radius than the unstable circular orbit corresponding to the maximum of the effective potential ($V''_{\rm eff} < 0$). 

The study of particle collisions and, in particular, the existence of an infinite center-of-mass energy during the collision, is directly linked to the existence of circular orbits, as was already pointed out previously (see e.g. \cite{Harada:2014vka}). In particular, the collision needs to happen ``close enough'' to the black hole horizon. Hence, to study particle collision near the horizon, we need to find the critical value of the angular momentum $L$ and its corresponding critical value of the radius $r$. The critical angular momentum is computed from the condition that a particle just barely reaches the event horizon of a black hole, i.e. it asymptotically approaches a radius outside the horizon with zero radial velocity $\dot{r}=0$. To determine the range of angular momentum values that allow the particle to approach the black hole, we must examine the effective potential via the conditions for the circular orbits given above. Furthermore, we require the orbits to be unstable in order to locate the turning point at which the particles change direction and fall into the horizon, which implies $V''_{\rm eff} <0$. This means that the conditions for the critical angular momentum are the same as those for an unstable circular orbit. The difference, however, is that we assume the colliding particles to be at rest at infinity, which fixes their energy. The conditions $V_{\rm eff}=V'_{\rm eff}=0$ give~:
\begin{equation}
\label{eq:collision}
(E +q V)^2\left(\frac{N'}{N} - \frac{2}{r} + 2 \frac{\sigma'}{\sigma}\right)  
-2 V' q( E + q V) + N' \sigma^2 (1-r^2) + 2r N\sigma^2 = 0 \ .
\end{equation}
In the following, we will choose $E$ such that the particle is at rest at infinity and calculate the value of $r=r_{c}$ at which the above equation is fulfilled. 
On of the interesting quantities in particle collisions is the center-of-mass energy $E_{\text{\tiny{C.M.}}}$. To evaluate this for the collisions we follow the approach introduced in \cite{Bardeen:1972}. From the coordinate basis vectors, we can form a tetrad which provides an orthonormal local inertial reference frame. To form a tetrad, we define an orthonormal vector in this coordinate system as~:
\begin{equation}
\textbf{e}_{m}=e_{m}{}^{\mu}\textbf{e}_{\mu} \ ,
\end{equation}
where \(e_{m}{}^{\mu}\) is a normalization associated with each basis vector. The Greek indices \((\mu,\nu)\) are associated with the coordinate basis, whereas Latin indices \((m,n)\) represent the axes of the tetrad frame. This allows us to choose local coordinates in which the metric \(g_{\mu\nu}\) becomes the flat Minkowski metric \(\eta_{mn}\) such that 
\[g_{\mu\nu}=\eta_{mn}e^{m}{}_{\mu}e^{n}{}_{\nu} \ .\]
In special relativity, the 4-velocity \(u^{\mu}\) is defined as the derivative of the space-time coordinates \(x^{\mu}=(t,x,y,z)\) with respect to proper time \(\tau\) and reads $u^{\mu}=\frac{dx^{\mu}}{d\tau}$.
The proper time is related to coordinate time \(t\) through the velocity \(\textbf{v}=(v_x,v_y,v_z)\) of the particle such that \(d\tau=dt/\gamma\) where \(\gamma=1/\sqrt{1-v^2}\) and \(v^2=v_x^2+v_y^2+v_z^2\). And so the 4-velocity in terms of \(\gamma\) and \(\textbf{v}\) is \(u^0=\gamma c\) and \(u^i=\gamma v_i\) for \(i=x,y,z\). Therefore, the 4-velocity is $u^{\mu}=\gamma(1,v_x,v_y,v_z)$.
The 4-momentum of a particle is defined as \(p^{\mu}=mu^{\mu}\) where \(m\) is its rest mass. From the definition of the 3-velocity, we know that \(v_i=u^i/u^0\)
which in the orthonormal frame becomes~:
\[v_{m}=\frac{e^m{}_{\mu}e^{\mu}}{e^m{}_{0}e^{\mu}} \ .\]
To study the center-of-mass energy of particle collisions, we begin by considering two particles moving towards a black hole which are initially at rest at infinity with rest masses \(m_1\) and \(m_2\) such that \(m_1\neq m_2\). The total momentum for the two particles in the tetrad frame is  $P^{m}=p^{m}_{(1)}+p^{m}_{(2)}=m_1u^{m}_{(1)}+m_2u^{m}_{(2)}$
and the center-of-mass energy is given by~:
\begin{align*}
E^2_{\text{\tiny{C.M.}}}=-P^{m}P_{m}&=m_1^2+m_2^2-2m_1m_2\eta_{mn}u^{m}_{(1)}u^{n}_{(2)}=m_1^2+m^2_2-2m_1m_2g_{\mu\nu}u^{\mu}_{(1)}u^{\nu}_{(2)} \ ,
\end{align*}
which for \(m_1=m_2=m_0\), i.e. two particles of equal mass, reduces to
\begin{equation}
\label{ECM}\frac{E^2_{\text{\tiny{C.M.}}}}{2m_0^2}=1-g_{\mu\nu}u^{\mu}_{(1)}u^{\nu}_{(2)} \ .
\end{equation}

The 4-velocities of two particles in the space-time given by the line element (\ref{eq:metric}) are~:
\begin{equation}
\label{4static}
u^{\mu}_{(i)}=\bigg(\frac{\tilde{E}_{i}}{N\sigma^2},\sqrt{\frac{\tilde{E}^2_i}{\sigma^2}-N\bigg(1+\frac{L^2_i}{r^2}\bigg)},0,\frac{L_i}{r^2}\bigg) \ , \ i=1,2 \ ,
\end{equation}
where $\tilde{E}_i=E_i + q_i V(r)$. 
Substituting (\ref{4static}) into (\ref{ECM}), we find the center-of-mass energy to be~:
\begin{equation}
\label{CMEstatic}
\frac{E^2_{\text{\tiny{C.M.}}}}{2m_0^2}=1+\frac{\tilde{E}_1 \tilde{E}_2}{N\sigma^2}-\frac{L_1L_2}{r^2}-\frac{1}{N}\sqrt{\frac{\tilde{E}_1^2}{\sigma^2}-N\bigg(1+\frac{L^2_1}{r^2}\bigg)}\sqrt{\frac{\tilde{E}^2_2}{\sigma^2}-N\bigg(1+\frac{L^2_2}{r^2}\bigg)} \ .
\end{equation}
Near the horizon, the expression for the center-of-mass energy is
\begin{equation}
\label{eq:com_energy}
\lim_{r\rightarrow r_h}\frac{E^2_{\text{\tiny{C.M.}}}}{2m_0^2}=1+\frac{\tilde{E}_1}{2\tilde{E}_2}\bigg(1+\frac{L^2_{2}}{r_h^2}\bigg)+\frac{\tilde{E}_2}{2\tilde{E}_1}\bigg(1+\frac{L^2_{1}}{r_h^2}\bigg)-\frac{L_{1}L_{2}}{r_h^2} \ .
\end{equation}
Note that this does not depend explicitly on the metric function $\sigma$, but that the value of $L$ at which $\dot{r}=0$ does (see (\ref{eq:lz2})). 
\section{The Reissner-Nordstr\"om limit}
In the limit where $\phi\equiv 0$, the solution of the field theoretical equations (\ref{eq:m}) - (\ref{eq:phi}) is the Reissner-Nordstr\"om (RN) solution (\ref{eq:RN}). The motion of (un)charged test particles in this space-time has been studied in \cite{Grunau:2011gd}. Here, we remind the reader of the main features of the ISCOs as well as test particle collisions (see also \cite{Zaslavskii:2010}). 

\subsection{Circular orbits of (un)charged particles}\label{sec:RN_nocharge}
In the following, we fix the black hole horizon at $r_h=1$, and hence have
\begin{equation}
	M=\frac{1}{2}r_h\left(1+\frac{K^2}{r_h^2}\right) \ \ , \ \ K^2:=\alpha(Q^2+Q_m^2) \ .
\end{equation}
We choose $K$ with $K\leq M$ to ensure a black hole solution with the extremal limit at $K=M$.
Circular orbits occur at stationary points of $r$, allowing us to find them analytically. The expression (\ref{eq:lz2}) gives for $q=0$, $\sigma\equiv 1$ and $N(r)=1-2M/r+K^2/r^2$~:
\begin{equation}\label{L_z}
	L^2=\frac{r^2\left(rM-K^2\right)}{r^2-3rM+2K^2} \ .
\end{equation}

This has solutions in $L$ for all $r>\frac{1}{2}\left(3M+\sqrt{9M^2-8K^2}\right)$.
There is for each $K$, a minimal value of $L$ for circular orbits. For each sufficiently large $L$ there exist two circular orbits~: one unstable orbit corresponding to a maximum of the effective potential, and one stable orbit corresponding to a minimum at a greater radius. As an example, let us discuss the extremal RN solution for $r_h=M=K=1$. In this case,
choosing e.g. $L=2$ will give an effective potential without stationary points and hence
no circular orbits.  In contrast, for $L=4$, the effective potential possesses a maximum and minimum. There is therefore an unstable circular orbit at the maximum located at $r=2.35$ and a stable circular orbit at the minimum located at $r=13.66$.

\begin{figure}[h!]
\begin{center}
\includegraphics[width=8cm]{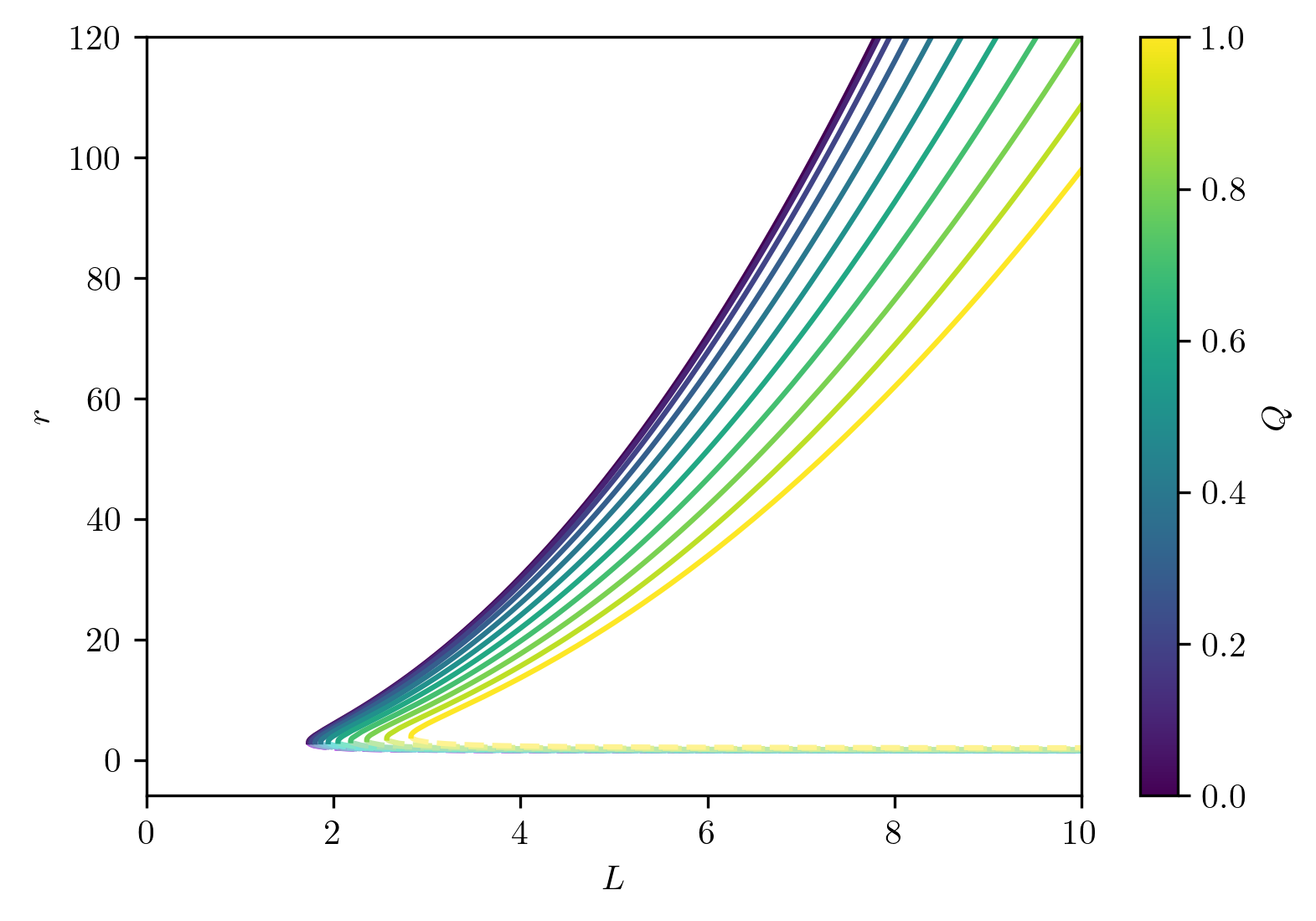}
\includegraphics[width=8cm]{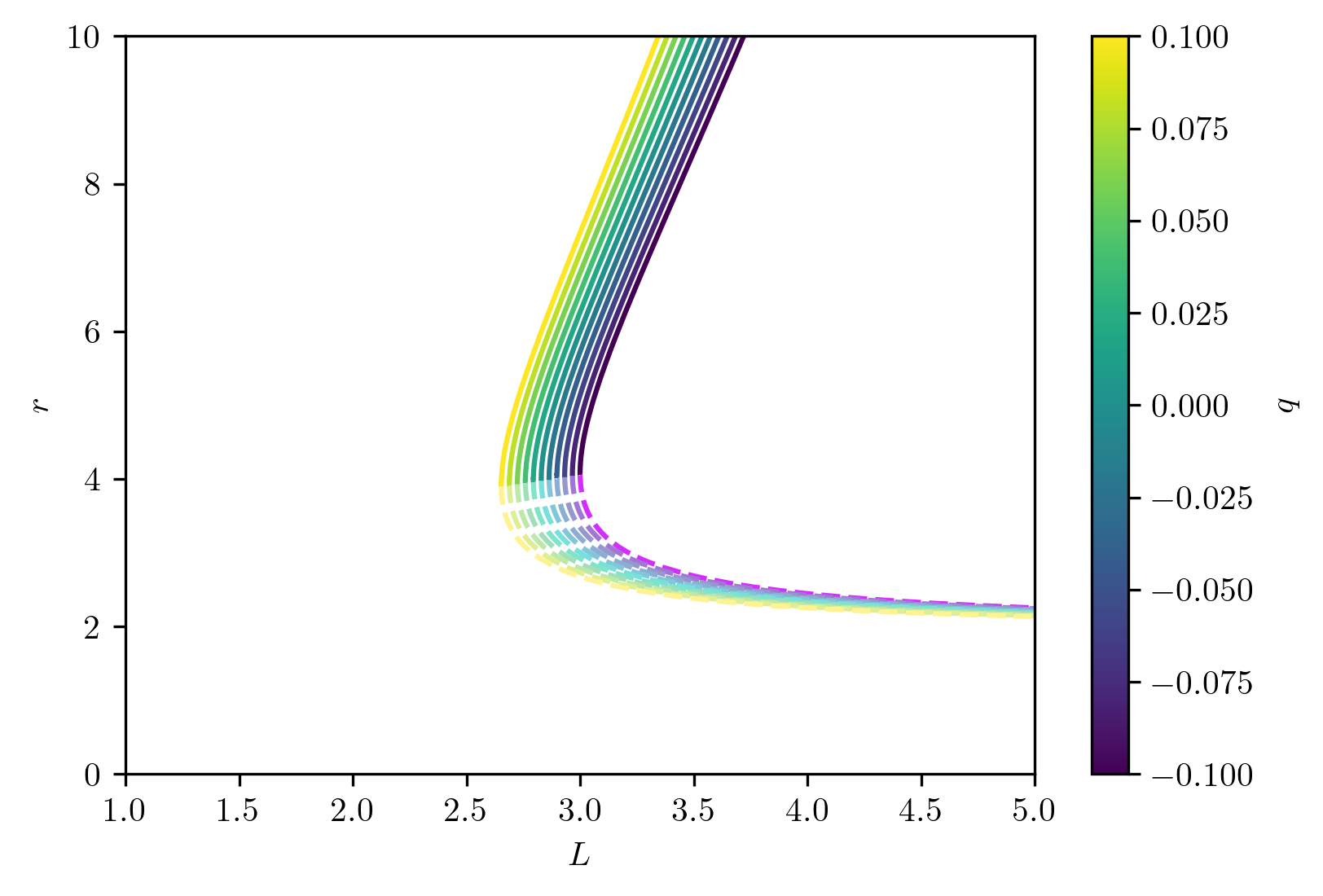}
\caption{{\it Left}: We show the radii of stable (solid) and unstable (dashed) circular orbits in the RN black hole space-time for different values of $Q$ in dependence on the angular momentum $L$ of uncharged test particles. {\it Right}: We show the radii of stable (solid) and unstable (dashed) circular orbits in the extremal RN black hole space-time $r_h=M=Q=1$ in dependence on the angular momentum $L$ of test particles with charge $q$.} 
\label{fig:RN_extremal_ISCO}
\end{center}
\end{figure}

In Fig.~\ref{fig:RN_extremal_ISCO} (left) we show the radius $r$ of the unstable circular orbit (dashed lines) and the radius of the stable circular orbit (solid lines) in dependence of the angular momentum $L$ in the space-time of a RN black hole for various of $Q$. The smallest possible radius of a stable circular orbit is the radius of the Innermost stable circular orbit (ISCO) which can be derived from the effective potential by letting $V_{\rm eff}=V'_{\rm eff}=V''_{\rm eff}=0$ and corresponds to the meeting point of the stable and unstable branch. These ISCOs were studied in detail in \cite{Pugliese:2010ps}.

Circular orbits for charged particles in the RN space-time were previously studied in detail \cite{Pugliese:2011py}. It was found that the radius of the ISCO is the solution to the third order equation
\begin{equation}
Mr^3 - 6 M^2 r^2 +9 M K^2 r- 4 K^2 = 0 \ .
\end{equation} 
The solutions to this are $r=6M$ for $K=0$ and $r=4M$ for $K=M$. In the following, we will remind the reader of the main features. 
Figure \ref{fig:RN_extremal_ISCO} (right) shows the radii of both stable (solid) and unstable (dashed) circular orbits for particles with different values of $q$ in the space-time of an extremal RN black hole with $M=K=r_h=1$. We observe that for particles with negative charge $q$, that is of opposite sign to the black hole charge $K$, the ISCOs have smaller radii than for the uncharged case. This is due to the electromagnetic attraction between the charge $q$ and the black hole. Equivalently, for positively charged particles ($q > 0$), the electromagnetic repulsion pushes the radii of the ISCOs to larger values. It is also clear that the ISCOs of negatively charged particles occur at greater angular momentum than for the uncharged case, with the opposite true for particles with positive charge. 

An interesting question in this context is whether test particles can remain static for an asymptotic observer in a given space-time. Since we have chosen $\dot{\theta}=0$ and $\dot{r}=0$ for the orbits discussed here,
this requires $\dot{\varphi}=0$, i.e. $L=0$. For uncharged particles ($q=0$) it is easy to see from (\ref{L_z}) that $L=0$ implies $rM-K^2=0$. For the static particle to be outside the horizon, we have to require $r > M$ with $r=M$ corresponding to the particle being located on the horizon of an extremal RN black hole. Hence, $rM = K^2 > M^2$ which cannot be fulfilled for a RN black hole. One can also show that charged particles ($q\neq 0$) cannot be static in the (electrically or dyonically) charged RN black hole space-time unless the particle resides on the horizon of an extremal RN black hole. The prove is given in Appendix \ref{Appendixb}. 

\subsection{Particle collisions for (un)charged particles}
\begin{figure}[h!]
\begin{center}
{\includegraphics[width=8cm]{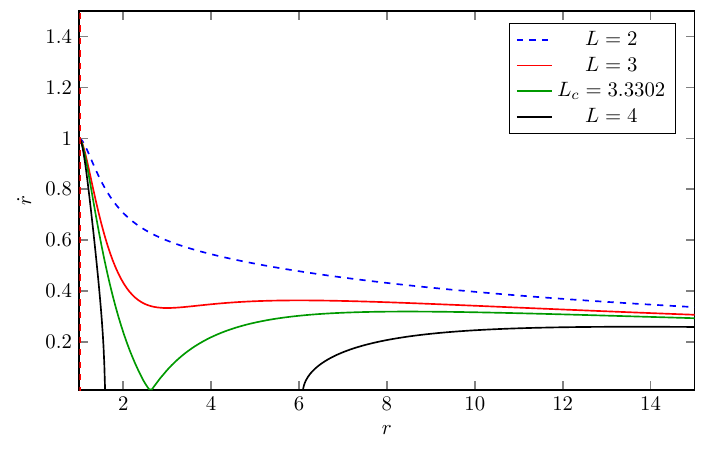}}
{\includegraphics[width=8cm]{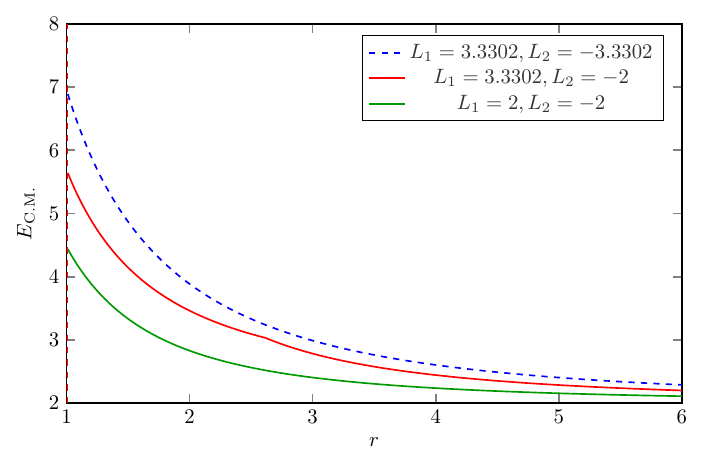}}
\vspace{-0.05cm}
\caption{{\it Left}: We show the radial velocity $\dot{r}$ in function of $r$ for uncharged particles in the extremal RN space-time with $r_h=M=K=1$ for several values of $L$. The $y$-axis corresponds to the horizon at $r=r_h=1$.
{\it Right}: We show the center-of-mass energy $E_{\text{\tiny{C.M.}}}$ in function of $r$ for three combinations of $L_{1}$ and $L_{2}$.  The $y$-axis corresponds to the horizon at $r=r_h=1$. }
\label{fig:RNcollisions}
\end{center}
\end{figure}

We first investigate the collisions of uncharged particles in the RN space-time, focusing specifically on their center-of-mass energy.  From $V_{\rm eff}=0$ we find for the RN case~:
\begin{equation}
\label{eq:RNlz2}
L^2=r^2\left(\frac{r^2(E^2-1)+2Mr-K^2}{r^2-2Mr+K^2}\right)  \ .
\end{equation}
From $V'_{\rm{eff}}=0$ we obtain an additional expression for $L^2$. Equating the latter with (\ref{eq:RNlz2}) we find the value of $r$ at which $V_{\rm eff}=V'_{\rm eff}=0$. This is a quartic equation in $r$~:
\begin{equation}
(E^2-1)r^4+(4M-3ME^2)r^3+(2K^2(E^2-1)-4M^2)r^2+4MK^2r-K^4=0  \ .
\end{equation}
We can fine-tune the balance between particle energy and black hole charge-mass ratio such that $M^2=K^2E^2$. This simplifies the quartic equation to
\begin{equation}
((E^2-1)r^2+Mr-K^2)(r^2-3Mr+K^2)=0
\end{equation}
and the following four roots are~:
\begin{displaymath}
r^{(\pm)}_1=\frac{-M\pm\sqrt{M^2+4K^2(E^2-1)}}{E^2-1}, \hspace{0.4cm} r^{[\pm)}_2=\frac{3M\pm\sqrt{9M^2-4K^2}}{2}  \ .
\end{displaymath}
As an example, let us discuss the extremal RN case. With our choice of $r_h=1$ this implies $K=M=1$ and corresponds to the particles having energy $E=1$.
This implies, that the only solution for $r > r_h$ (we want the collision to happen outside the horizon) in this case is $r=r_2^{(+)}=(3+\sqrt{5})/2$. Inserting this into (\ref{eq:RNlz2}) we find $L\approx 3.3302$. Note that this choice also satisfies $V''_{\text{eff}}<0$ as required. In Fig. \ref{fig:RNcollisions} (left), we show the radial velocity $\dot{r}$ in function of $r$ for different choices of $L$. \mc{Since we focus on infalling particles, for which \(\dot{r}<0\), we follow the convention adopted in the literature and plot \(|\dot{r}|=\dot{r}\).} 
For $L=4$, $\dot{r}$ becomes zero indicating a turning point. However, the particle cannot reach the horizon at $r_h=1$. For $L=3.3302$ the radial velocity $\dot{r}$ becomes zero close to the horizon and particles can reach the horizon. For smaller values of $L$ all geodesics fall in, but do not possess a point at which $\dot{r}=0$.

To compute the center-of-mass energy we first write the 4-velocities of two uncharged particles moving in the RN space-time~:
\begin{equation}
u^{\mu}_{(i)}=\bigg(\frac{E_ir^2}{r^2-2Mr+K^2},\sqrt{E_i^2-\bigg(\frac{r^2-2Mr+K^2}{r^2}\bigg)\bigg(1+\frac{L^2_{i}}{r^2}\bigg)},0,\frac{L_{i}}{r^2}\bigg) \ , \ \ i=1,2 \ .
\end{equation}
Inserting this into (\ref{ECM}) gives the center-of-mass energy~:
\begin{align*}
\label{CMEstatic}
\frac{E^2_{\text{\tiny{C.M.}}}}{2m_0^2}&=1+\frac{E_1E_2r^2}{r^2-2Mr+K^2}-\frac{L_{1}L_{2}}{r^2}\\&-\frac{r^2}{r^2-2Mr+K^2}\sqrt{E_1^2-\bigg(\frac{r^2-2Mr+K^2}{r^2}\bigg)\bigg(1+\frac{L^2_{1}}{r^2}\bigg)}\sqrt{E^2_2-\bigg(\frac{r^2-2Mr+K^2}{r^2}\bigg)\bigg(1+\frac{L^2_{2}}{r^2}\bigg)} \ \ ,
\end{align*}
which near the horizon becomes~:
\begin{equation}
\label{CME2.1}
\lim_{r\rightarrow r_h}\frac{E^2_{\text{\tiny{C.M.}}}}{2m_0^2}=1+\frac{E_1}{2E_2}\bigg(1+\frac{L^2_{2}}{r_h^2}\bigg)+\frac{E_2}{2E_1}\bigg(1+\frac{L^2_{1}}{r_h^2}\bigg)-\frac{L_{1}L_{2}}{r_h^2} \ \ .
\end{equation}
As a result, the center-of-mass energy will always be finite. In Fig. \ref{fig:RNcollisions} (right) we show $E_{\text{\tiny{C.M.}}}$ in function of $r$ for several combinations of $L_{1}$ and $L_{2}$.

Let us now turn to the collision of charged particles. The collisions of these particles 
have been discussed previously \cite{Zaslavskii:2010}, however in most cases, the angular momentum $L$ has been set to zero implying that the particle moves only radially. We will study the more general case and allow the partilce to possess angular momentum.
From (\ref{eq:lz2}) we find~:
\begin{equation}
L^2=\frac{r^2}{2\Sigma^2}\bigg(qQ^2(\Delta-\Sigma)-2\Sigma\Delta\pm r\sqrt{qQ(\Delta-\Sigma)(4\Sigma\Delta+qQ(\Delta-\Sigma)r^2+4\Sigma^2qQ/r^2)}\bigg) \ ,
\end{equation}
where $\Sigma=3Mr-r^2-2K^2$ and $\Delta=Mr-K^2$. To determine the value of $E$ for which $r$ becomes a turning point, we observe that~:
\begin{equation}
E_{\pm}=\pm\sqrt{\bigg(1-\frac{2M}{r}+\frac{K^2}{r^2}\bigg)\bigg(1+\frac{L^2}{r^2}\bigg)}-qV(r)  \ .
\end{equation}
An energy value carrying a positive (negative) sign is associated with the solution characterized by~:
\begin{displaymath}
    \lim_{r\rightarrow\infty} E_{\pm}=\pm 1-qV_{\infty} \ ,
    \end{displaymath}
such that \(E_+(L, q, r)\geq E_-(L_, q, r)\) with the following relation \(E_+(L,q,r)=-E_-(L,-q,r)\) \cite{Pugliese:2011py}. 

The 4-velocity of electrically charged particles in the RN space-time are~:
\begin{equation}
u^{\mu}_{(i)}=\bigg(\frac{(E_i+q_iV(r))r^2}{r^2-2Mr+K^2},\sqrt{(E_i+q_iV(r))^2-\bigg(\frac{r^2-2Mr+K^2}{r^2}\bigg)\bigg(1+\frac{L^2_i}{r^2}\bigg)},0,\frac{L_i}{r^2}\bigg) \ , \ \ i=1,2 \ .
\end{equation}
Since the trajectories are timelike, $\dot{t}\geq 0$ reduces to the condition 
$E_i\geq 0$ on the horizon where $V(r_h)=0$ due to our choice of gauge for the field theoretical solutions. Since the energy is conserved, we find that $Q\leq \pm 1/q $. The center-of-mass energy is~:
\begin{equation}
\begin{split}
\frac{E^2_{\text{\tiny{C.M.}}}}{2m_0^2}
&=1+\frac{\big(E_1+q_1V(r)\big)\big(E_2+q_2V(r)\big)}{N_{\rm RN}}
-\frac{L_1L_2}{r^2}\\
&\quad-\frac{1}{N_{\rm RN}}
\sqrt{\big(E_1+q_1V(r)\big)^2-N_{\rm RN}\bigg(1+\frac{L_1^2}{r^2}\bigg)}
\sqrt{\big(E_2+q_2V(r)\big)^2-N_{\rm RN}\bigg(1+\frac{L_2^2}{r^2}\bigg)} \ \ , 
\end{split}
\end{equation}
where $N_{\rm RN}=1-2M/r + K^2/r^2$. Near the horizon this gives~:
\begin{align}
\lim_{r\rightarrow r_h}\frac{E^2_{\text{\tiny{C.M.}}}}{2m_0^2}&=1+\frac{1}{2}\bigg(\frac{E_1+q_1V(r_h)}{E_2+q_2V(r_h)}\bigg)\bigg(1+\frac{L^2_2}{r_h^2}\bigg)+\frac{1}{2}\bigg(\frac{E_2+q_2V(r_h)}{E_1+q_1V(r_h)}\bigg)\bigg(1+\frac{L^2_1}{r_h^2}\bigg)-\frac{L_1L_2}{r_h^2} \ ,
\end{align}
which becomes infinite for $E_i+q_i V(r_h)=0$ for either $i=1$ or $i=2$. With $E_i=1-q_i V_{\infty}$ this gives
$q_i=Q/r_h$ which is independent of the angular momentum. Let us compare this to the $L=0$ case discussed in \cite{Zaslavskii:2010}. Different to our case, where we have chosen the gauge such that $V(r_h)=0$, this is not used in   \cite{Zaslavskii:2010}. Hence, the critical point is at $r=r_h$ and consequently $E=Qq/r_h$. However, if the particles has $q=0$, this is unphysical as this would imply $E=0$, i.e. we need to require the particle collision to happen at $r > r_h$. The same result was obtained by \cite{Kimura:2011} for radially infalling particles, one into an extremely charged  black hole where one of the particles had no charge, and the other had $E=q$. This suggests that in spherically symmetric backgrounds, we can obtain infinite center-of-mass energies for arbitrary angular momentum \cite{Zaslavskii:2010}. In fact, this is the difference to the original case of particles colliding in a Kerr space-time \cite{Banados:2009}. In this latter case, neutral particles falling into an extremal Kerr black hole were considered. One of the particles is a critical particle in the sense that its angular momentum is fine-tuned so that it can orbit arbitrarily close to the horizon without immediately plunging in. The other particle is generic and a divergent center-of-mass energy arises when a critical particle collides near the horizon with a generic particle. The angular momentum fine-tuning makes the particle's local radial velocity vanish near the horizon. This is analogous to the effective energy near the horizon going to zero. However, when a charged particle falls into an extremal RN, the divergence occurs because $E+qV(r_h)$ vanishes.  The value of $L=0$ is necessary since a circular orbit at the horizon is not possible. This is due to an electrostatic potential that cancels the energy, producing {\it critical} particles in a radial fall, not orbit. In the Kerr case \cite{Banados:2009} the critical particle orbits the horizon at zero radial velocity and the collision with a generic particle produces large relative velocity, which in return gives infinite center-of-mass energy. In the RN space-time, the particle's energy is canceled by the potential near the horizon, causing it to radially fall into the black hole, producing infinite center-of-mass energy when it collides with a generic particle. 

\begin{figure}[h!]
\begin{center}
{\includegraphics[width=7cm]{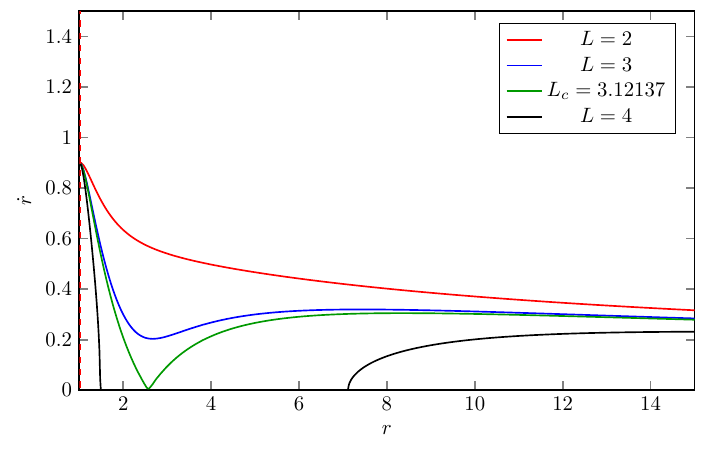}}
{\includegraphics[width=7cm]{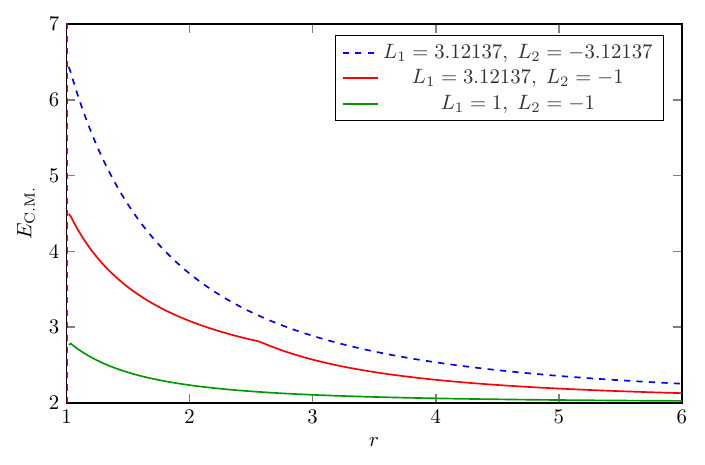}}
\vspace{-0.05cm}
\caption{{\it Left}: We show the radial velocity $\dot{r}$ in function of $r$ for a charged particle with $q=0.1$ in the extremal RN space-time with $r_h=M=K=1$ for different values of $L$. $r=r_h=1$ correspond to the extremal horizon. The critical angular momentum is $L=3.12137$.
{\it Right}: We show the center-of-mass energy $E_{\text{\tiny{C.M.}}}$ in function of $r$ for three combinations of $L_{1}$ and $L_{2}$ for a charged particle with $q=0.1$ in the extremal RN space-time with $r_h=M=K=1$. $r=r_h=1$ corresponds to the extremal horizon. 
} 
\label{fig:RNcharged1}
\end{center}
\end{figure}

\begin{figure}[h!]
\begin{center}
{\includegraphics[width=7cm]{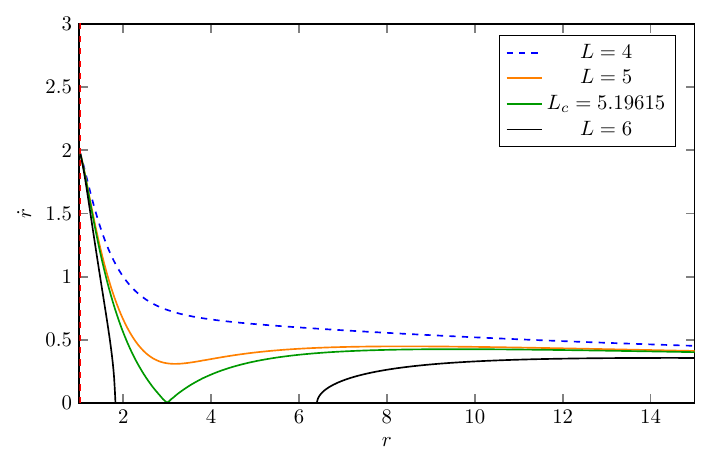}}
{\includegraphics[width=7cm]{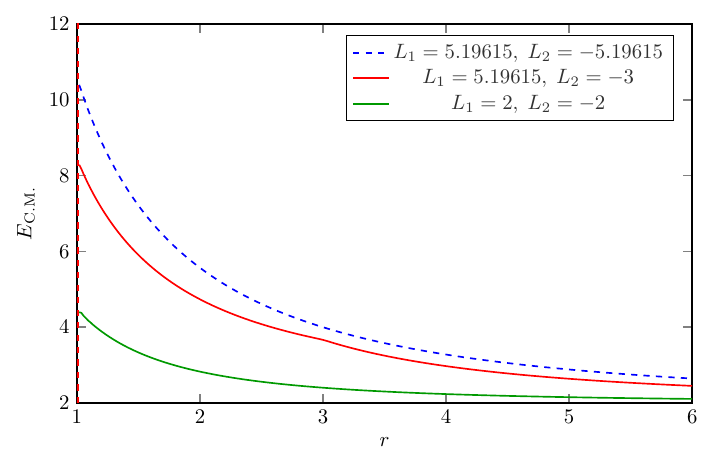}}
\vspace{-0.05cm}
\caption{{\it Left}: We show the radial velocity $\dot{r}$ in function of $r$ for a charged particle with $q=-0.1$ in the extremal RN space-time with $r_h=M=K=1$ for different values of $L$. $r=r_h=1$ corresponds to the extremal horizon. The critical angular momentum is $L=5.19615$.
{\it Right}: We show the center-of-mass energy $E_{\text{\tiny{C.M.}}}$ in function of $r$ for three combinations of $L_{1}$ and $L_{2}$ for a charged particle with $q=-0.1$ in the extremal RN space-time with $r_h=M=K=1$. $r=r_h=1$ corresponds to the extremal horizon. }
\label{fig:RNcharged2}
\end{center}
\end{figure}
In Fig. \ref{fig:RNcharged1} (left), we show the radial velocity $\dot{r}$ of a charged particle with $q=0.1$ moving in an extremal RN space-time for several values of $L$. For $L=3.12137$ we find a critical point ($\dot{r}=0$) at $r=2.57238$. In Fig.\ref{fig:RNcharged1} (right) we show the center-of-mass energy $E_{\text{\tiny{C.M.}}}$ in dependence of $r$ for three combinations of $L_{1}$ and $L_{2}$. We find that the center-of-mass energy is finite in all cases. 
In Fig. \ref{fig:RNcharged2} we show the corresponding results for a negatively charged particle $q=-0.1$. We find that $\dot{r}=0$ at $r=3$ for $L=5.19615$ (see left) and that the center-of-mass is always finite (see right).

\begin{figure}[!h]
    \centering
    \includegraphics[width=0.5\linewidth]{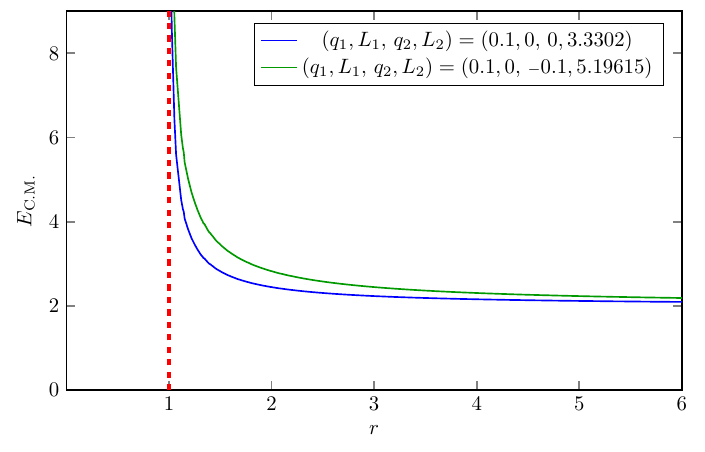}
    \caption{We show the center-of-mass energy in function of $r$ for a two particle collision in an extremal RN space-time with $r_h=M=K$. $L_{i}$ and $q_i$, $i=1,2$, denote the angular momentum and charge of the two particles.}
    \label{fig:RNcharged3}
\end{figure}
In these two cases, we see how the particle charge influences the value of the critical angular momentum and the radius at which $\dot{r}$. The center-of-mass energy is always finite, however, it is possible to have infinite center-of-mass energy for suitable choices of the charges and angular momenta. As mentioned above, one of the particles needs to have vanishing angular momentum, i.e. $L=0$. We show the center-of-mass energy as function of $r$ for two combinations in Fig.\ref{fig:RNcharged3}. In the cases for the collision of uncharged particle with a positively charged particle (blue line) and of two charged particles with opposite charge (green line) we see the center-of-mass energy diverge at the horizon of the black hole.

\section{Black holes with scalar hair}
In the following, we will investigate circular orbits and particle collisions in the space-time of charged black holes that carry scalar hair, i.e. the solutions discussed in Section \ref{sec:solutions}. We will first discuss our results for the electrically charged black holes ($k=0$), and then discuss the influence of the additional magnetic charge by studying dyonic black holes ($k=1$). We have interpolated the numerical solutions of the field equations (\ref{eq:m}) - (\ref{eq:phi}) using cubic splines in Python with a mesh of $50,000$ as well as with the Interpolation-command in MATHEMATICA. Both tools give us the same results (within numerical accuracy).

\subsection{Circular orbits}
We have studied both uncharged as well as charged particles in circular motion around charged black holes with scalar hair. We will first discuss the electrically charged black hole space-times and then compare with the dyonically charged case. 

\subsubsection{$k=0$: Electrically charged black holes with scalar hair}
\begin{figure}[!h]
	\centering
	\begin{subfigure}{0.4\textwidth}
		\includegraphics[width=\textwidth]{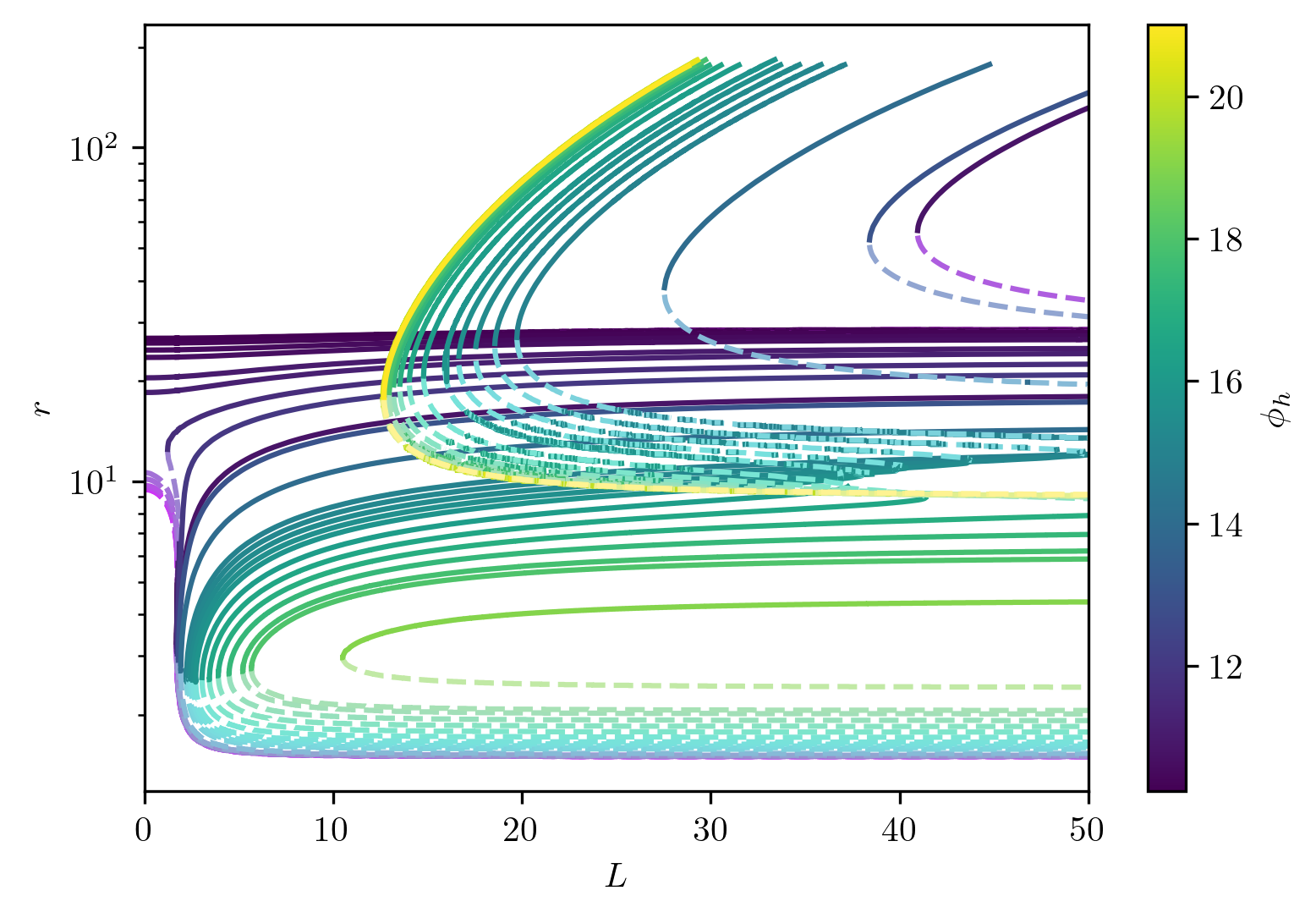}
	\end{subfigure}
	\begin{subfigure}{0.4\textwidth}
		\includegraphics[width=\textwidth]{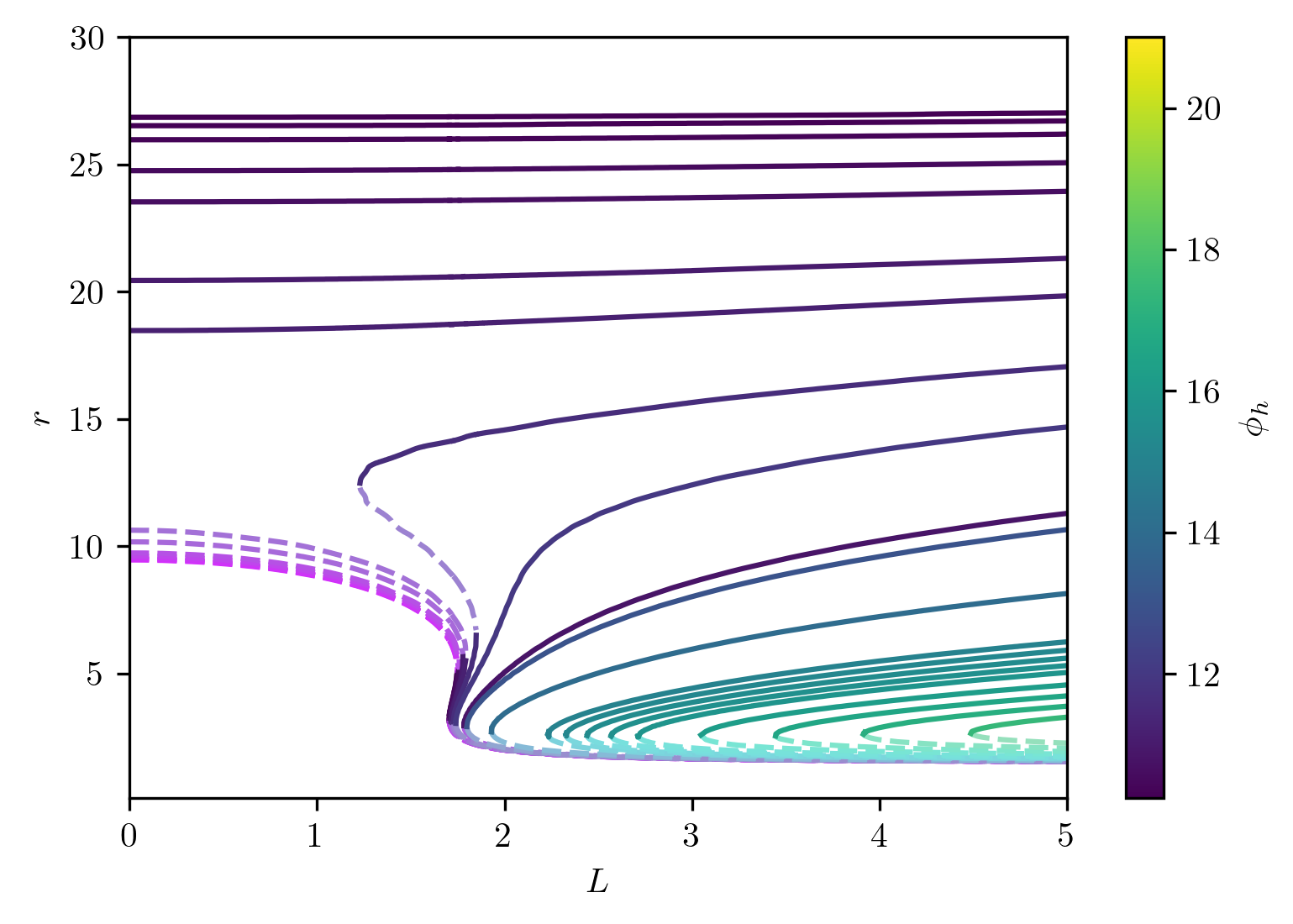}
	\end{subfigure}
	\caption{{\it Left:} We show the radii of stable (solid) and unstable (dashed) circular orbits of uncharged particles in dependence of the angular momentum $L$ in the space-time of an electrically charged black hole that carries scalar hair. Here $g=0.008$, $k=0$, $\alpha=0.001$ and we give the orbits for different values of the scalar field value on the horizon, $\phi(r_h)=\phi_h$. {\it Right:} Same as left, but zoomed into the interval $L\in [0:5]$.
	\label{fig:circular_uncharged_k_0_g_0_008}}
\end{figure}

For $g=0.008$ we note that both the ADM mass $M$ as well as the electric charge $Q$ have qualitatively different branches in $\phi_h$ (see Fig. \ref{fig:M_Q_vs_phirh}) which suggests that new phenomena might appear. Our results for the circular orbits of uncharged particles are shown in Fig. \ref{fig:circular_uncharged_k_0_g_0_008}.

We observe that for sufficiently large values of $\phi(r_h)$, the space-time essentially behaves like Reissner-Nordst\"om with no circular orbits for small $L$, but a stable and an unstable circular orbit for large $L$. Decreasing the value of $\phi(r_h)$, we observe differences to the RN case. This change happens at $\phi(r_h)\approx 14$. We find that for a specific range of angular momentum $L$ an additional pair of circular orbits exists. This can be seen in Fig.\ref{fig:circular_uncharged_k_0_g_0_008} (right). This range of $L$ increases when decreasing $\phi(r_h)$ further. At $\phi(r_h)\lesssim 12$ we observe a phenomenon not present in the RN case:
circular orbits can exist for $L=0$. These orbits correspond to static orbits (since $\dot{r}=\dot{\theta}=\dot{\varphi}=0$) and have been discussed first in rotating vacuum space-times \cite{Collodel:2017end} and consequently in static, spherically symmetric space-times \cite{Wei:2023bgp} with scalar and electromagnetic fields. In the latter case, it was pointed out that non-linear electrodynamics is necessary for the static orbits to exist. {\it Here, we demonstrate that static  orbits also exist in a standard Einstein-Maxwell-scalar field black hole space-time.} In agreement with the results in \cite{Wei:2023bgp} we find that one stable and one unstable static orbit exists. 

\begin{figure}[!h]
	\centering
	\begin{subfigure}{0.4\textwidth}
		\includegraphics[width=\textwidth]{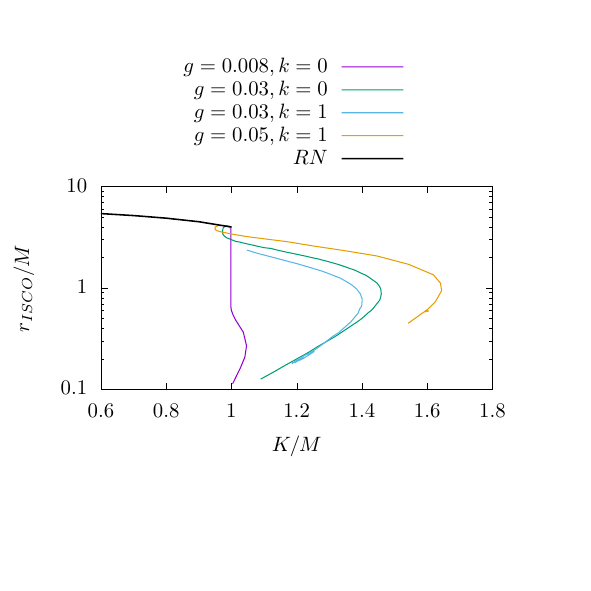}
	\end{subfigure}
	\begin{subfigure}{0.4\textwidth}
		\includegraphics[width=\textwidth]{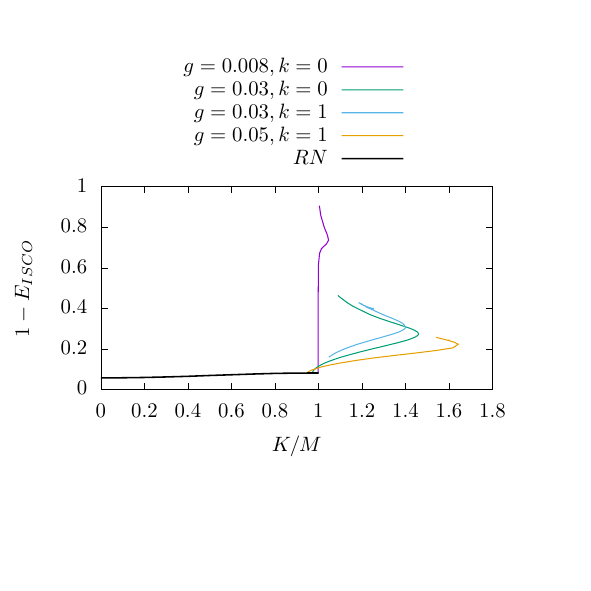}
	\end{subfigure}
	\vspace{-1cm}
	\caption{{\it Left}: We show the radius $r_{ISCO}$ of the ISCO in units of the black hole mass $M$
	in function of the mass to charge ratio $K/M$ for different values of $g$ and electrically charge ($k=0$) as well as dyonically charged ($k=1$) black hole solutions that carry scalar hair. For comparison, we also show the corresponding values for the RN solution.  {\it Right}: The efficiency for gravitational energy converted into radiation as the uncharged particle falls from infinity into the black hole
${\cal{E}}= E_{\infty} - E_{\rm ISCO}=1-E_{ISCO}$ in function of the mass to charge ratio $K/M$ for the same solutions.   
	\label{fig:rISCO_EISCO_all}}
\end{figure}

The innermost stable circular orbit (ISCO) corresponds to the point at which the stable and unstable branch of circular orbits meet. We find that both the radius of this orbit and the value of the angular momentum of the particle on this orbit increase with increasing $\phi(r_h)$.
In Fig.~\ref{fig:rISCO_EISCO_all} (left) we show the radius of the ISCO $r_{ISCO}$ in function of the charge to mass ratio $K/M$ of the solutions. We find that $r_{ISCO}/M$ is always smaller than the corresponding value for the RN black hole and that the branch of electrically charged black holes ($k=0$) with scalar hair for $g=0.008$ meets the branch of RN solutions for large values of $\phi(r_h)$. It also tends back to $K/M \rightarrow 1$ for small values of $r_{ISCO}/M$. 
This is related to the fact that in the limit of small enough $\phi(r_h)$ (we find $\phi(r_h)\approx 10.8$ to be the minimal value for the scalar field on the horizon) the solution develops a {\it hard wall} as described in \cite{Brihaye:2025nfx}. In this limit, the exterior solution becomes very close to the extremal RN solution. Hence, we find that when decreasing the value of the scalar field on the horizon the ratio $r_{ISCO}/M$ decreases. One important difference to the RN case is that
for a given value of $K/M$ we find two different radii for the ISCO.  

The energy $E=E_{\rm ISCO}$ of the particle on the ISCO gives 
the amount of gravitational energy converted into radiation as a particle falls down from infinity all the way to the black hole. The efficiency for this is given by
${\cal{E}}= E_{\infty} - E_{\rm ISCO}$, where $E_{\infty}$ is the energy of the particle at infinity which is equal to unity for uncharged particles. We find that for all charged black hole solutions, this conversion is more efficient than in the RN case with efficiency increasing when the value of the scalar field is decreasing reaching the highest efficiency in the limit of the ``hard wall` solution discussed above. 

\begin{figure}[!h]
    \centering
    \includegraphics[width=0.4\linewidth]{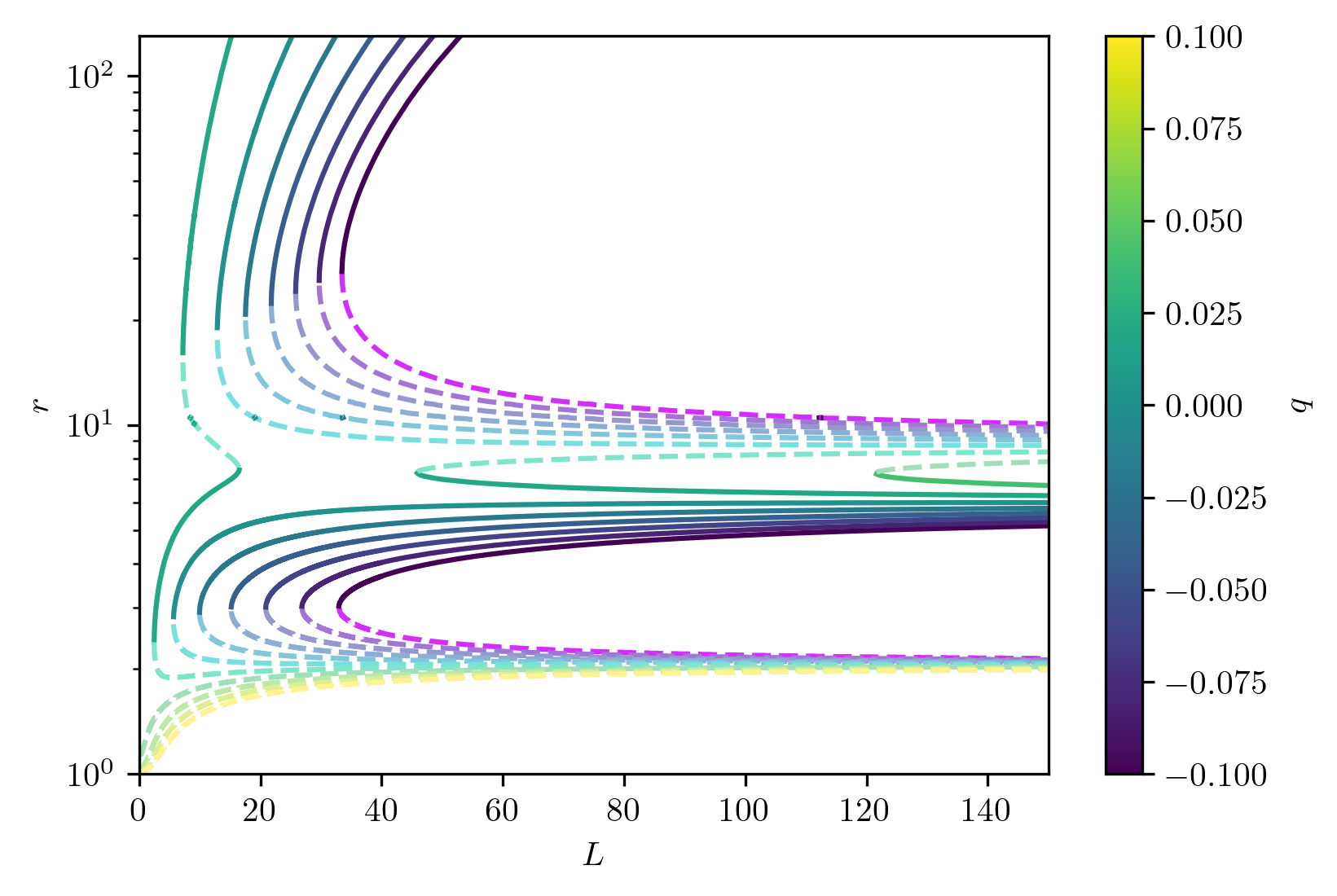}
    \includegraphics[width=0.4\linewidth]{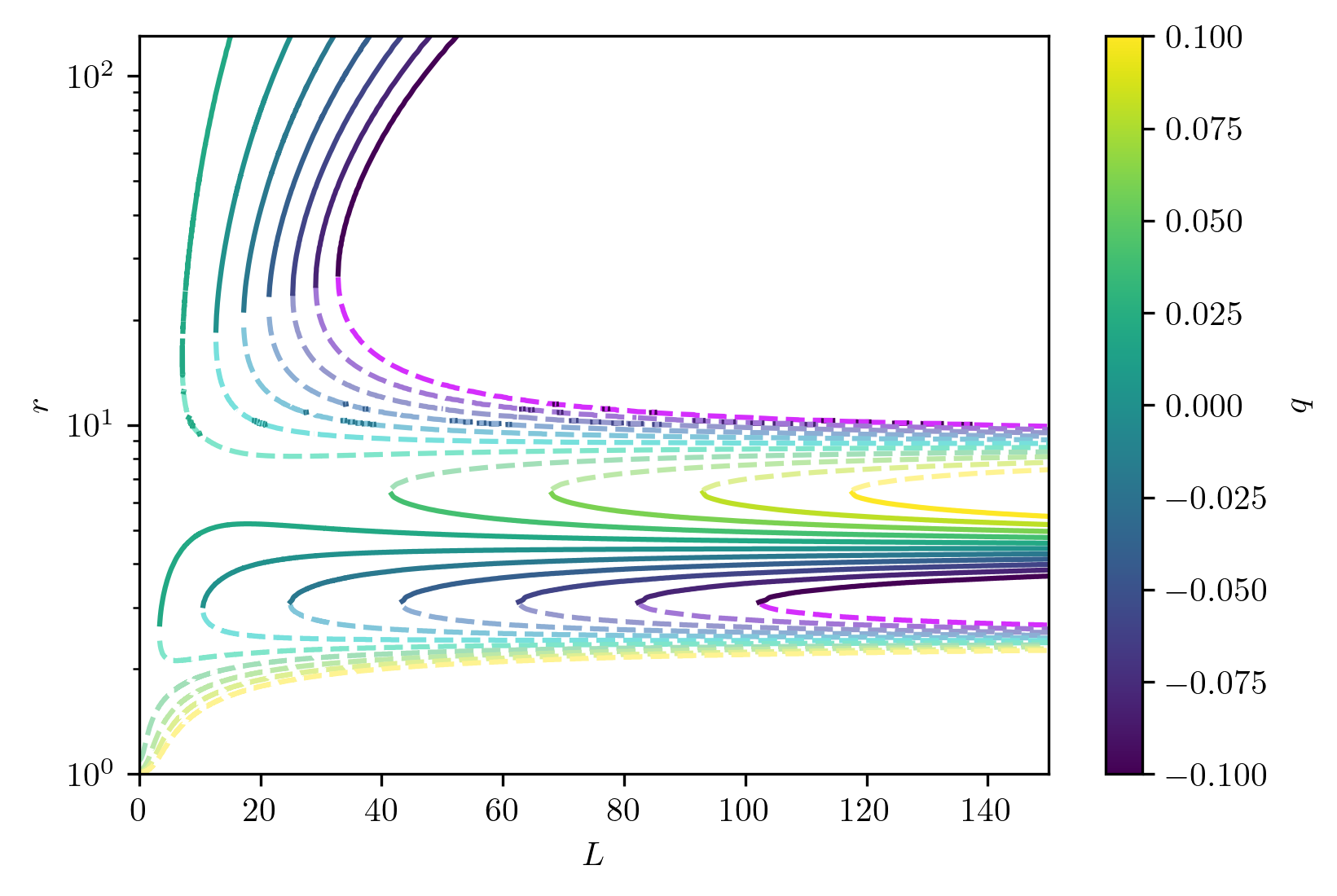}
    \caption{We show the radii of stable (solid) and unstable (dashed) circular orbits of charged particles in dependence of the angular momentum $L$ in the
    space-time of an electrically charged ($k=0$) black hole with scalar hair for $\alpha=0.001$ and $g=0.008$. The value of the scalar field on the horizon is $\phi(r_h)=\phi_h=18$ (left) and
  $\phi(r_h)=\phi_h=19$ (right), respectively.}
    \label{fig:circular_charged_g_0_008_k_0_ph_18}
\end{figure}

Next, we have studied the circular orbits for charged particles. In Fig.~\ref{fig:circular_charged_g_0_008_k_0_ph_18} we show the circular orbits for
charged particles in the space-time of an electrically charged black hole ($k=0$) with scalar field value on the horizon $\phi(r_h)=\phi_h=18$ (left) and
$\phi(r_h)=\phi_h=19$ (right), respectively. We see that for the $\phi_h=18$ case, negatively charged and neutral particles have two pairs of circular orbits for nearly all $L$. For particles at the ISCO of the outer branch of orbits, a small loss of angular momentum would therefore lead to a significant change in $r$. For particles with charge $q \geq 0.04$, we see the same single unstable orbit at small $L$ as in other sets of solutions, followed by the appearance of an unstable-stable pair at higher $L$. For $q=0.02$, however, we see the intermediate case with two pairs of orbits for small $L$ but only one for larger $L$. 
For $\phi_h=19$, while the outer pair of orbits shows very similar results, the ISCOs for negatively charged particles occur at similar $r$ but much larger $L$, with an increasing difference as the magnitude of the charge increases. For positively charged particles, in contrast, the ISCO occurs at much lower $L$, although again similar $r$. 
We therefore observe very different results when changing $\phi_h$ from $\phi_h=18$ to $\phi_h=19$, which are not immediately apparent in the neutral case but which will become relevant when charged particles are involved.

\begin{figure}[!h]
    \centering
    \includegraphics[width=0.4\linewidth]{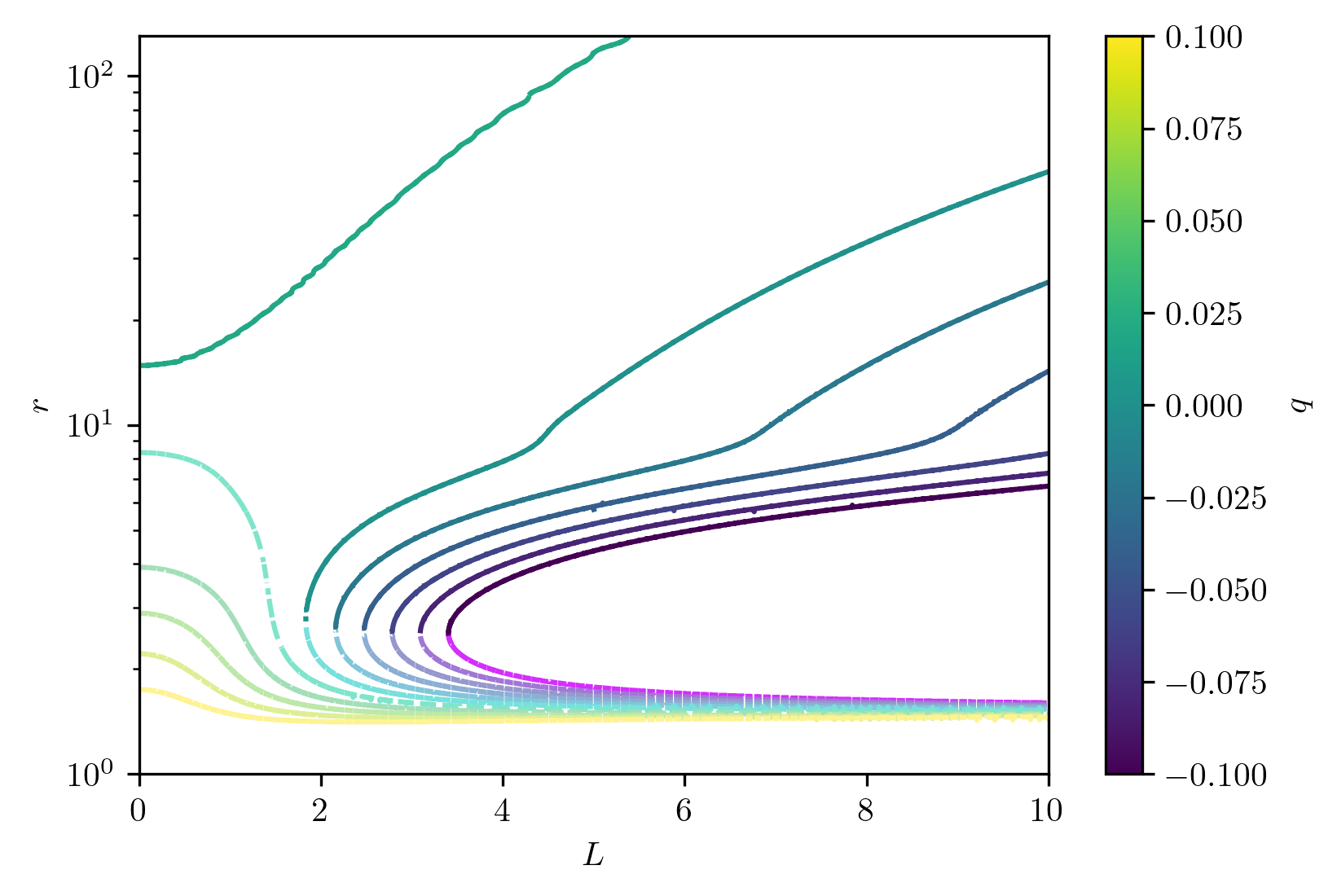}
    \includegraphics[width=0.4\linewidth]{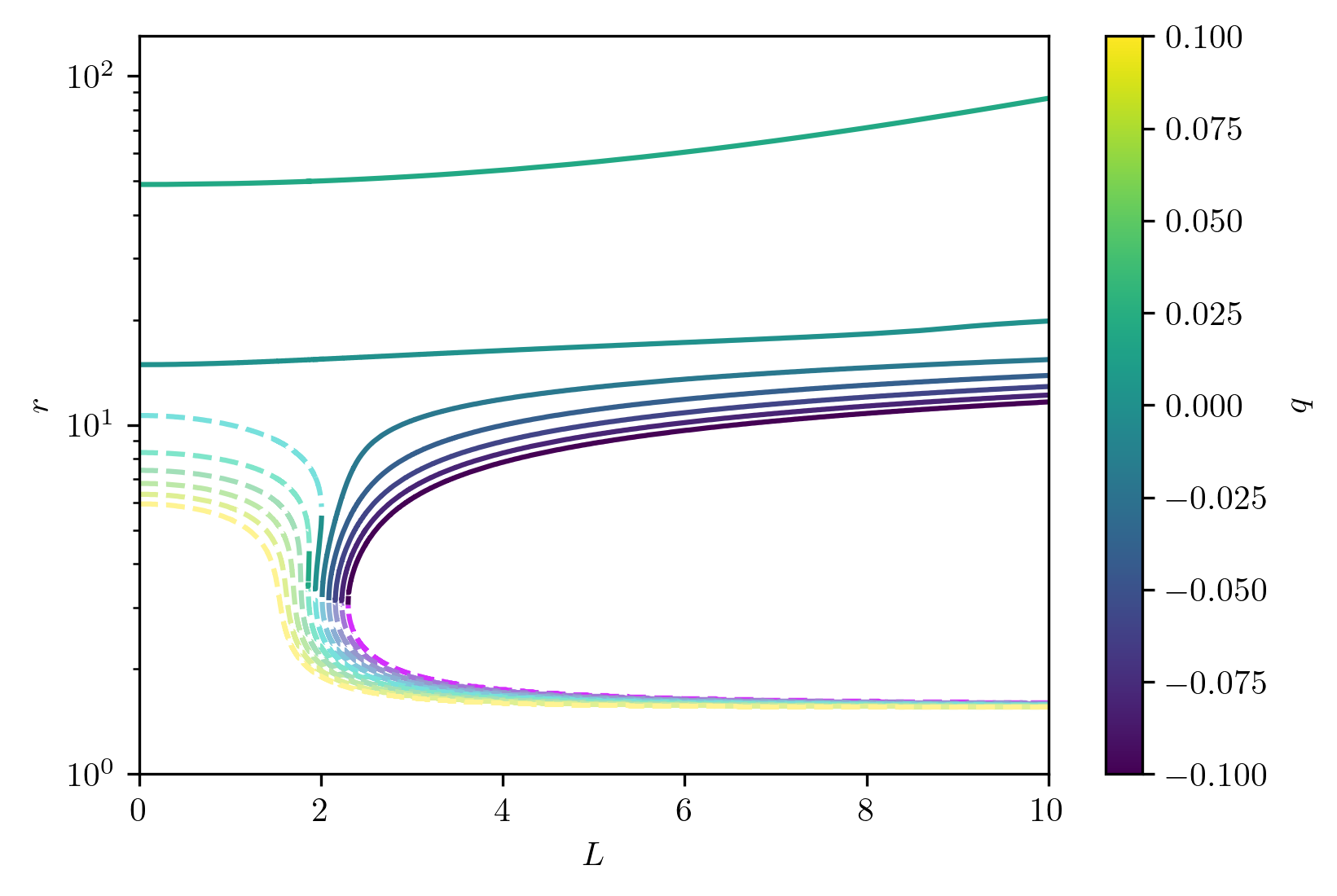}
    \caption{We show the radii of stable (solid) and unstable (dashed) circular orbits of (un)charged particles in dependence of the angular momentum $L$ in the
    space-time of a black hole with scalar hair for $\alpha=0.001$, $g=0.03$. On the left, we show the results for $k=0$ and $\phi(r_h)=\phi_h=5.6$, on the right we show the results for $k=1$ and $\phi(r_h)=\phi_h=1.5$, respectively. }
    \label{fig:circular_charged_g_0_03_k_0_ph_5_6}
\end{figure}

To understand the influence of the gauge coupling $g$ on the existence of circular orbits, we have also studied the case $k=0$ and $g=0.03$. Our results for the innermost pair of circular orbits for $\phi(r_h)=\phi_h=5.6$ are shown in Fig. \ref{fig:circular_charged_g_0_03_k_0_ph_5_6} (left).
Decreasing $q$ from zero, we find that the behaviour is very similar to the $q=0$ case: the radius of the ISCO is slightly decreasing. Since the particle is subject to an additional electromagnetic attraction, it needs to have larger angular momentum in order to be able to stay on the circular orbit, hence the value of $L$ on the ISCO increases with decreasing and negative $q$. Increasing $q$ from zero, we find that for sufficiently large $q$, static orbits with $L=0$ are possible. These orbits are unstable and we do not find 
stable circular orbits for sufficiently large and positive $q$.  

\begin{figure}[h!]
\begin{center}
{\includegraphics[width=8cm]{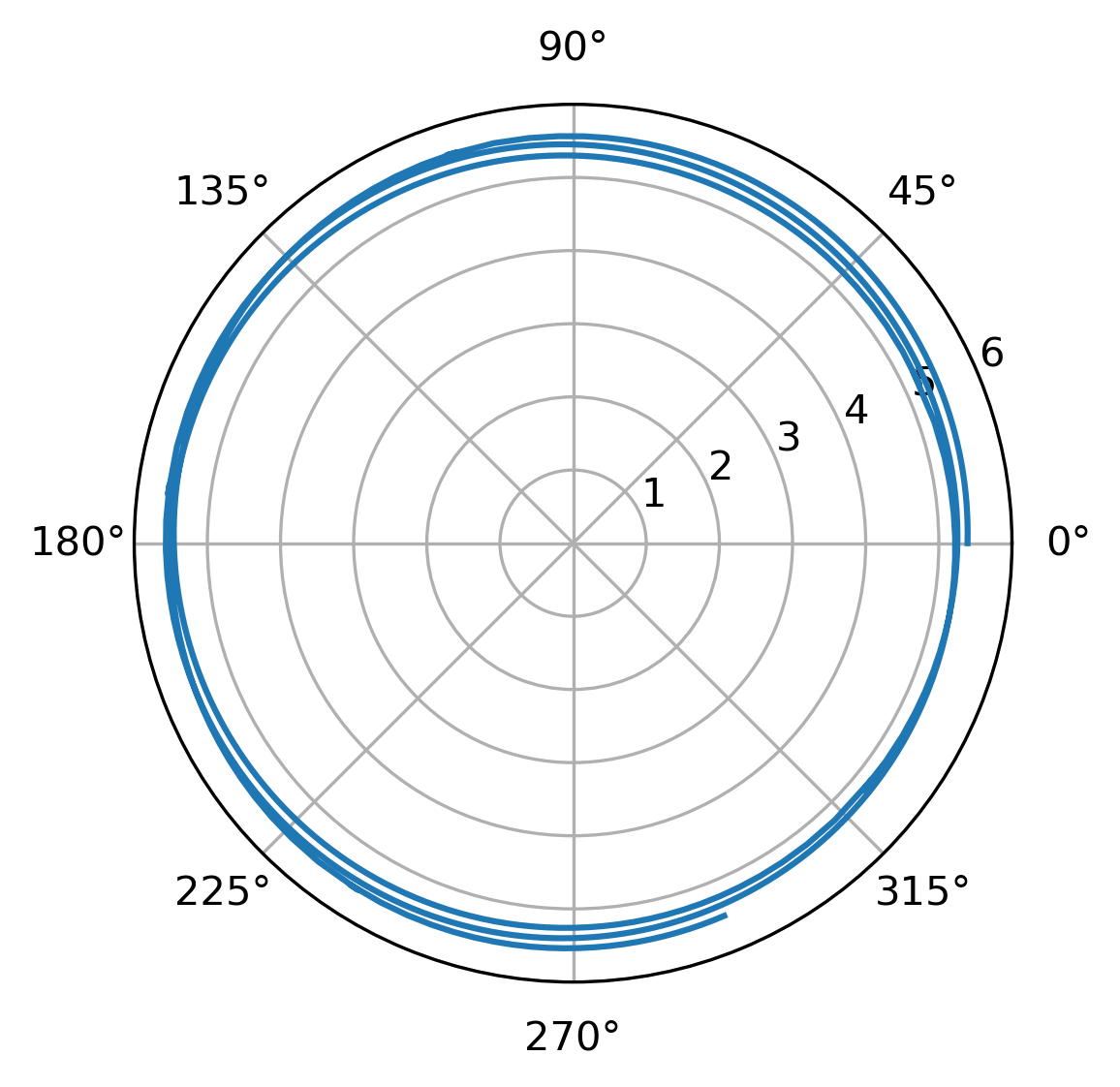}}
{\includegraphics[width=8cm]{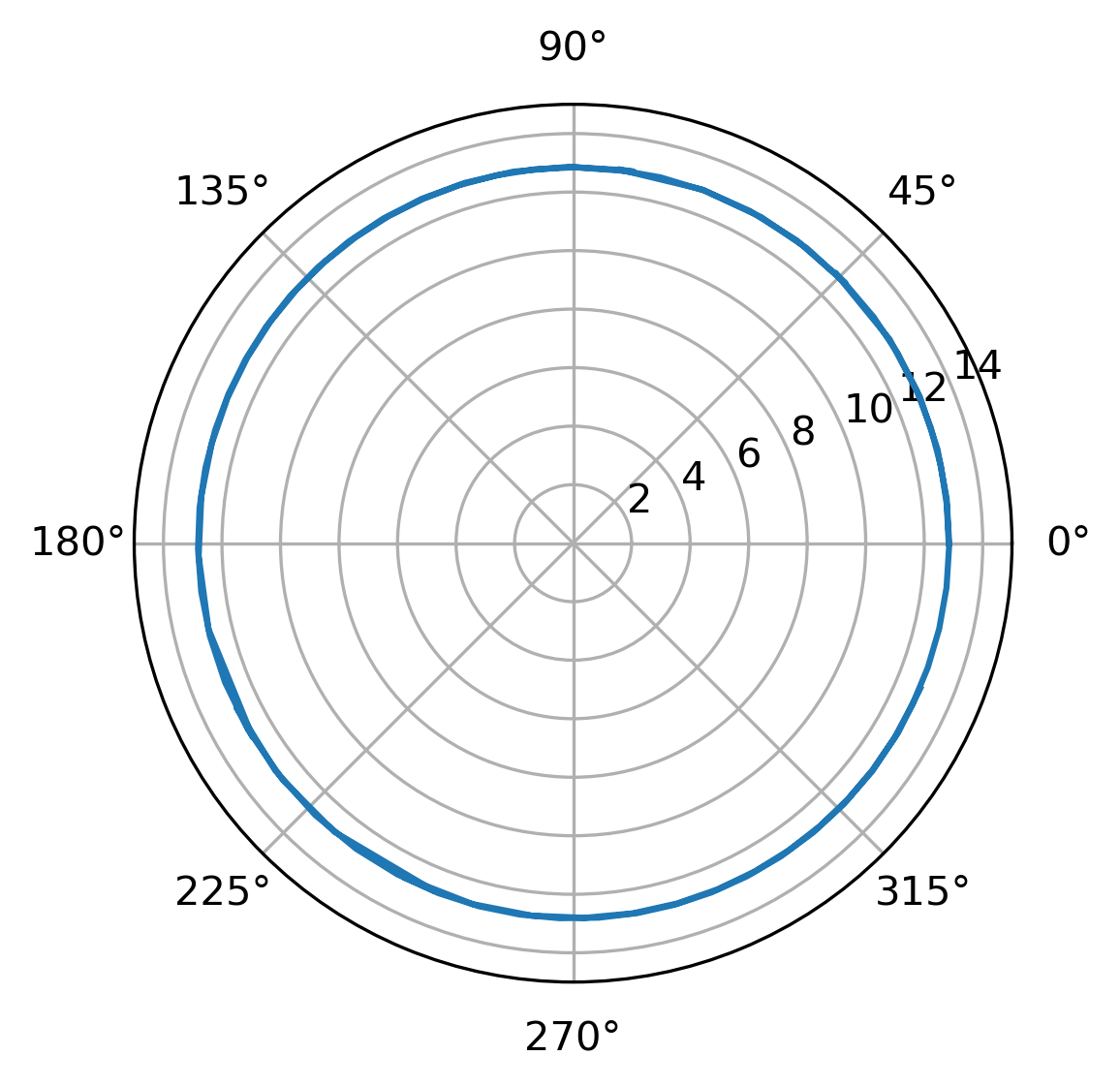}}
\caption{We show examples of stable (nearly) circular orbits of uncharged particles for $g=0.008$, $\phi_h=17$, $L=10$
(left) and $g=0.03$, $\phi_h=5.6$, $L=4$ (right) in the space-time of an electrically charged black hole ($k=0$) that carries scalar hair. For both cases $\alpha=0.001$ and $r_h=1.0$.
\label{fig:examples_circular}}
\end{center}
\end{figure}

In Fig.~\ref{fig:rISCO_EISCO_all} (left) we give $r_{ISCO}/M$ of uncharged particles for the $k=0$, $g=0.03$ solutions.
We find that, very similar to the $k=0$, $g=0.008$ case, the value of $r_{ISCO}/M$ is decreasing along the branch when decreasing the value of $\phi(r_h)$. At the largest possible value of $\phi(r_h)$, the branch joins the RN branch at a non-extremal solution. Again, we find that for a given value of $K/M$ two different ISCOs are present. For the values of $K/M$ for which solutions with different $g$ exist, the solution with $g=0.008$ has the smaller value of $r_{ISCO}/M$ as compared to the $g=0.03$ case. Moreover, the efficiency of gravitational energy conversion is higher in the $g=0.008$ case as compared to the $g=0.03$ case, see 
Fig.~\ref{fig:rISCO_EISCO_all} (right). We conclude that the radius of a circular orbit of an uncharged particle around an electrically charged black hole with scalar hair is always smaller than that in the corresponding RN case and that $r_{ISCO}/M$ decreases when choosing smaller values of the gauge coupling $g$. Examples of stable (nearly) circular orbits for both values of $g$ are given in   
Fig.~\ref{fig:examples_circular}. Note that the numerical treatment of the problem makes it extremely difficult to find the exact circular orbit.

\begin{figure}[h!]
\begin{center}
{\includegraphics[width=8cm]{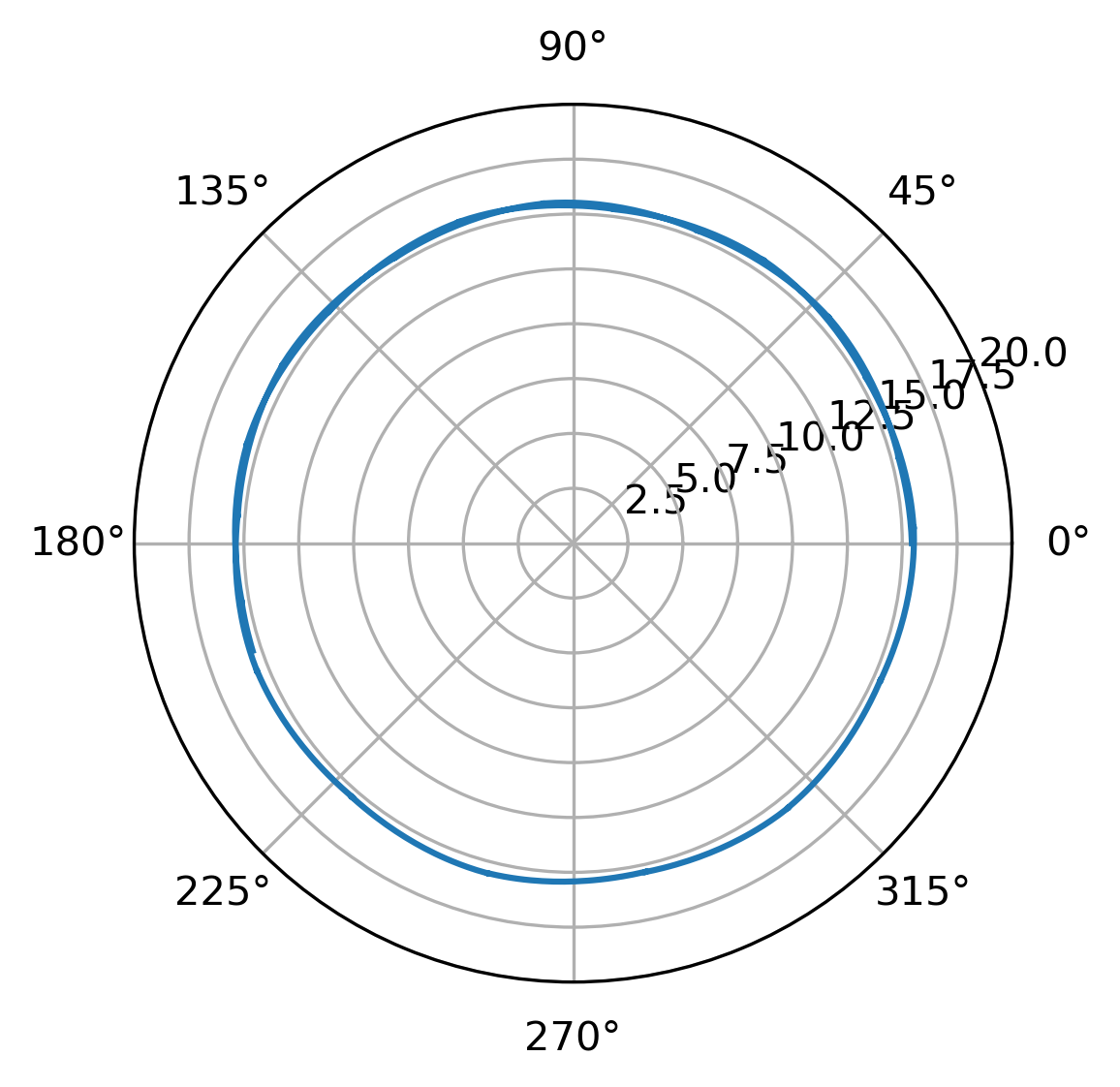}}
{\includegraphics[width=8cm]{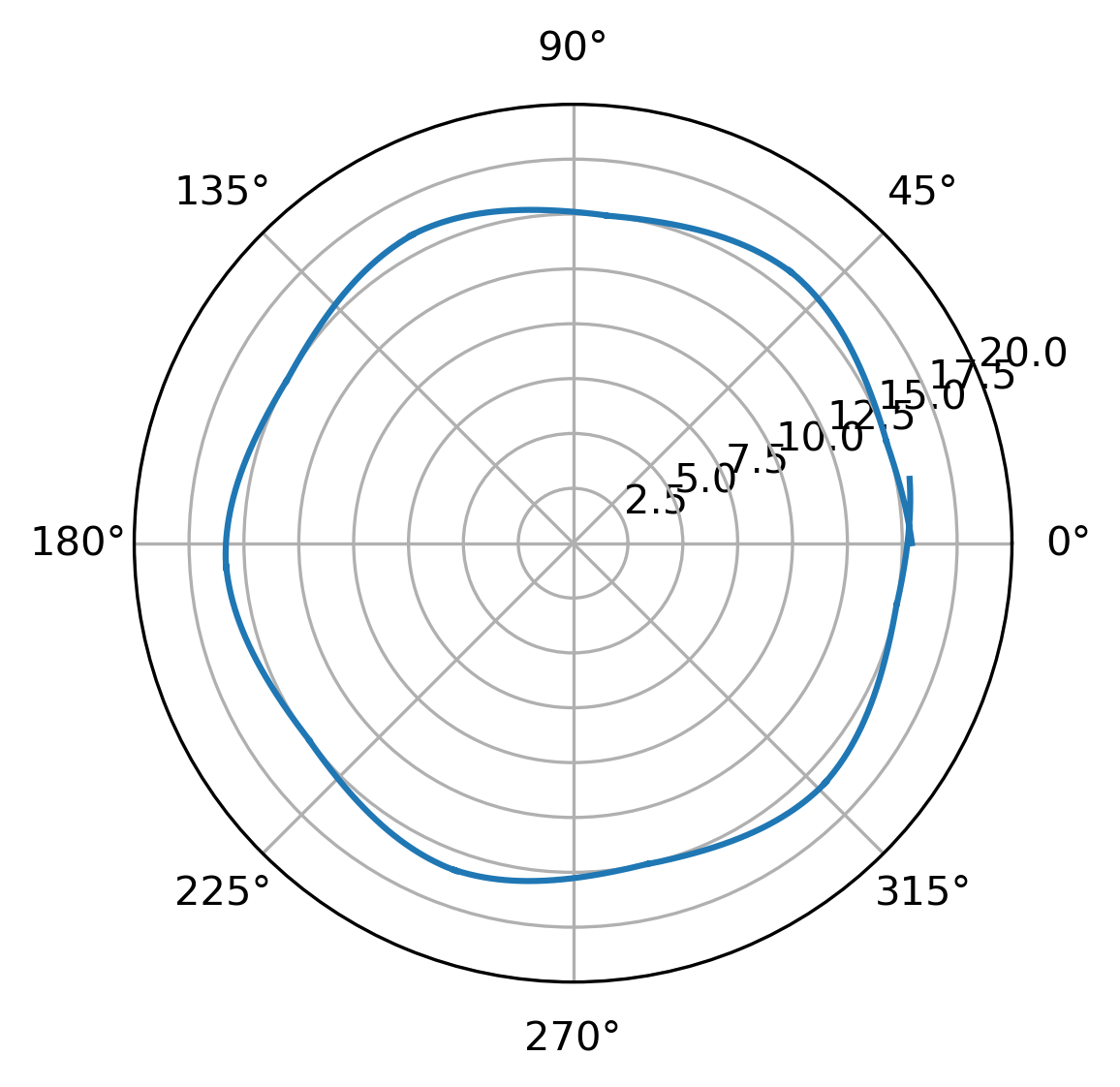}}
{\includegraphics[width=8cm]{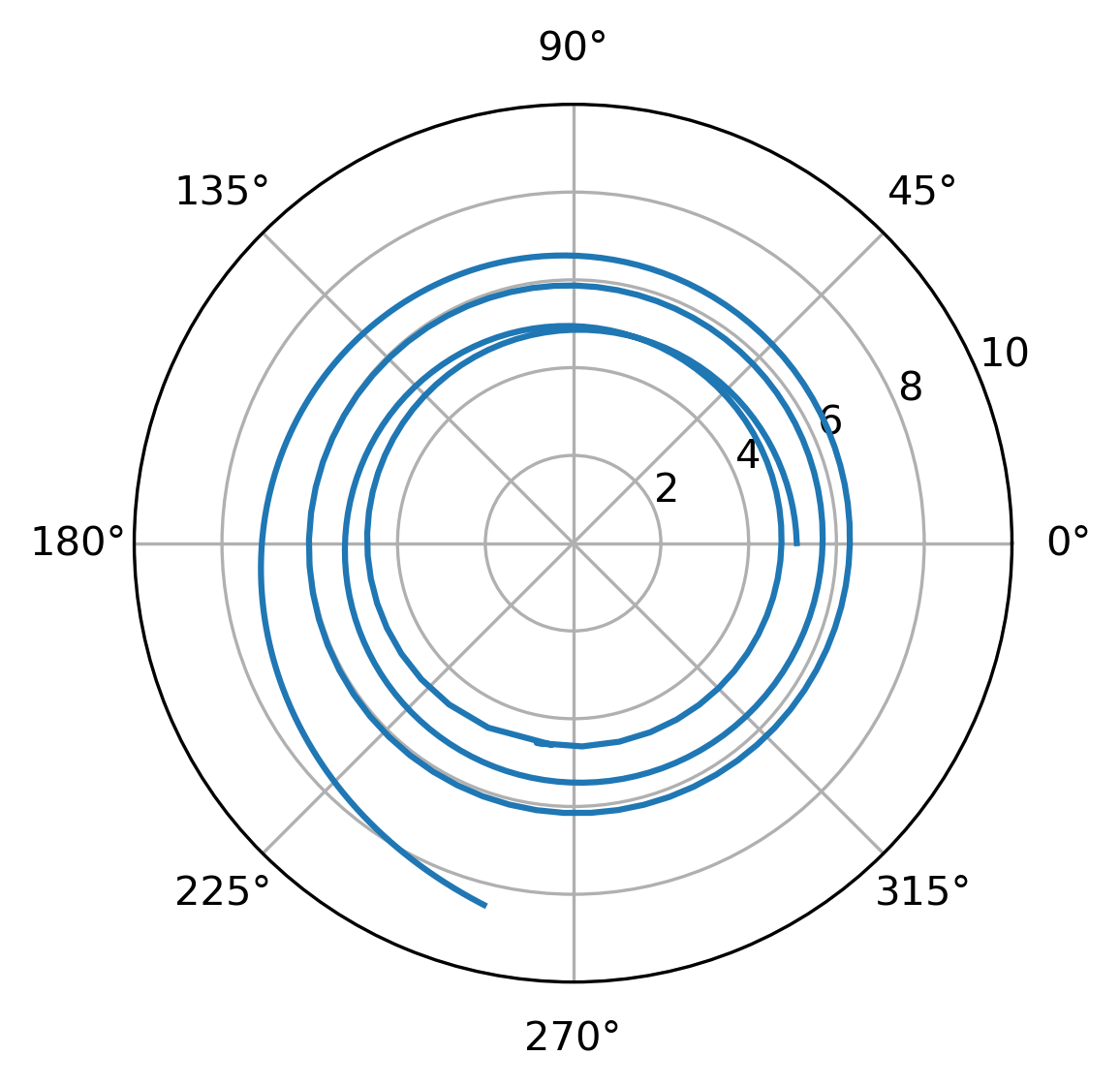}}
{\includegraphics[width=8cm]{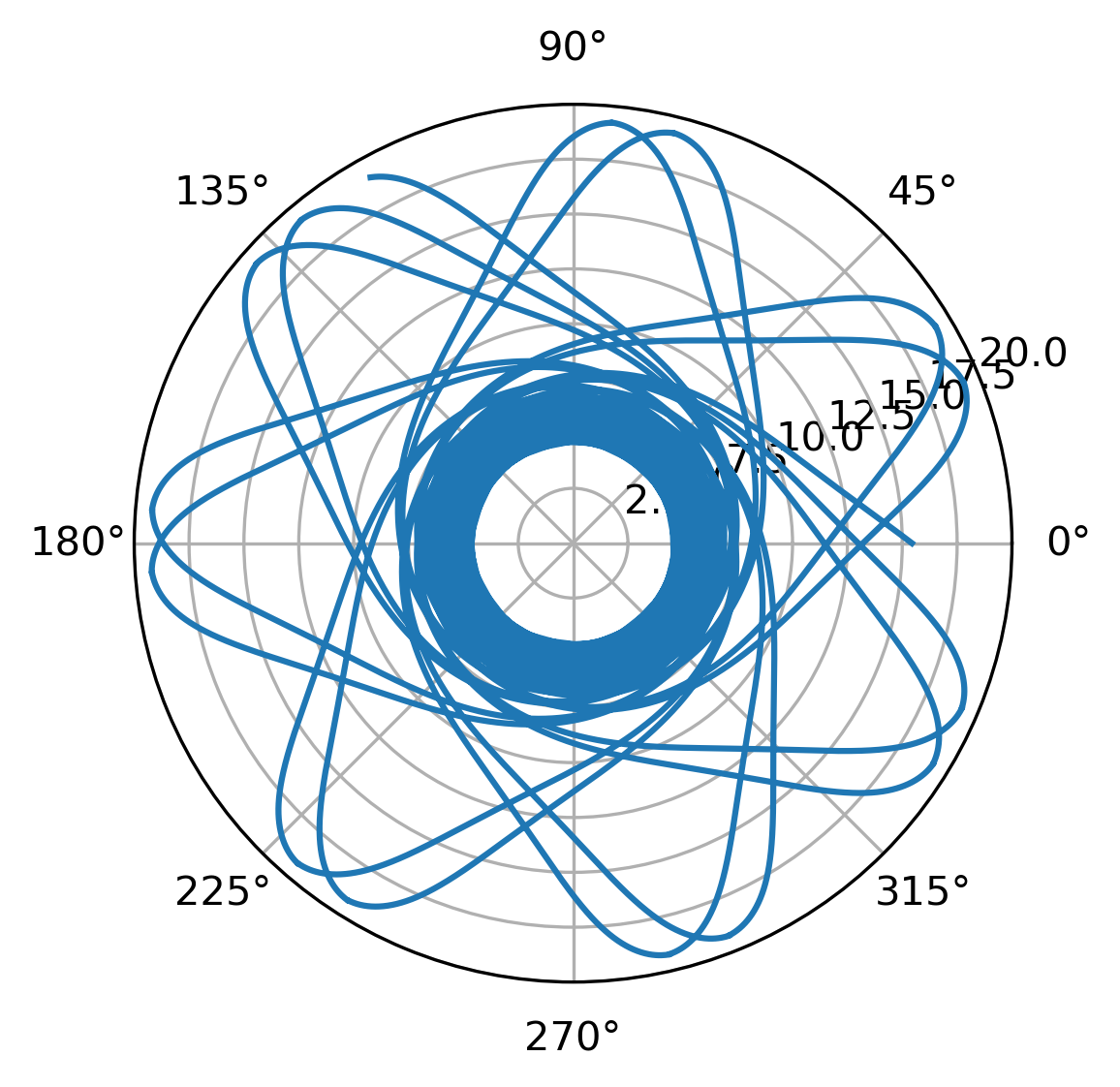}}
\caption{Examples of nearly-circular orbits for uncharged particle in the space-time of a dyonically charged black hole with scalar hair ($g=0.03, k=1, \phi_h=1.5$, $r_h=1$, $\alpha=0.001$). All orbits have $L=2$ and have their initial position very close to the minium of the effective potential. For the orbits in the upper row, the initial position is very close to the global minimum of the effective potential at $r=15.438428999659994$. Moreover $E=0.62807$ (upper left) and 
$E=0.6282$ (upper right), respectively. In the lower row, we show the orbit for a particle initially close to the smaller minimum at $r=5.0966867824336495$ and with energy $E= 0.6411$. On the left, the first few orbits close to the circular orbit, on the right the pattern forming over time.
\label{fig:examples_nearly_circular}
}
\end{center}
\end{figure}

\subsubsection{$k=1$: Dyonically charged black holes with scalar hair}

For uncharged particles $q=0$, we have first investigated the case $g=0.03$ in order to be able to compare with the $k=0$ case and understand the influence of the additional magnetic charge of the black hole. For uncharged particles ($q=0$) we find that for sufficiently low value of $\phi(r_h)$ we can find static orbits with $L=0$. For larger values of $\phi(r_h)$, the dependence of the radii of stable and unstable orbits is qualitatively similar to that of the RN black hole. We also find additional pairs of circular orbits. Hence, two stable circular orbits exist for specific parameter ranges corresponding to two local minima of the effective potential. This is similar to the $g=0.008$, $k=0$ case. In order to demonstrate how difficult it is to find the exact circular orbit, we show an example of the two stable circular orbits for $\phi_h=1.5$ in Fig.\ref{fig:examples_nearly_circular}. The radius of the global minimum of the effective potential corresponds to $r=15.438428999659994$ and energy $E=0.62807$. As the figure
shows, changing the energy in its fourth digit, i.e. choosing $E=0.6282$ changes the (nearly) circular orbit to an orbit with slight oscillations in its radius. For the local minimum of the potential corresponding to the second stable circular orbit at $r=5.0966867824336495$ with energy
$E=0.6411$, which is slightlyy off the value for the ``real'' circular orbit. As can be seen, the orbit is initially (nearly) circular, but moves away from the near-circularity over time. We hence conclude that the global minimum seems to be ``more stable'' with respect to slight variations in energy.

\begin{figure}[!h]
    \centering
    \includegraphics[width=0.4\linewidth]{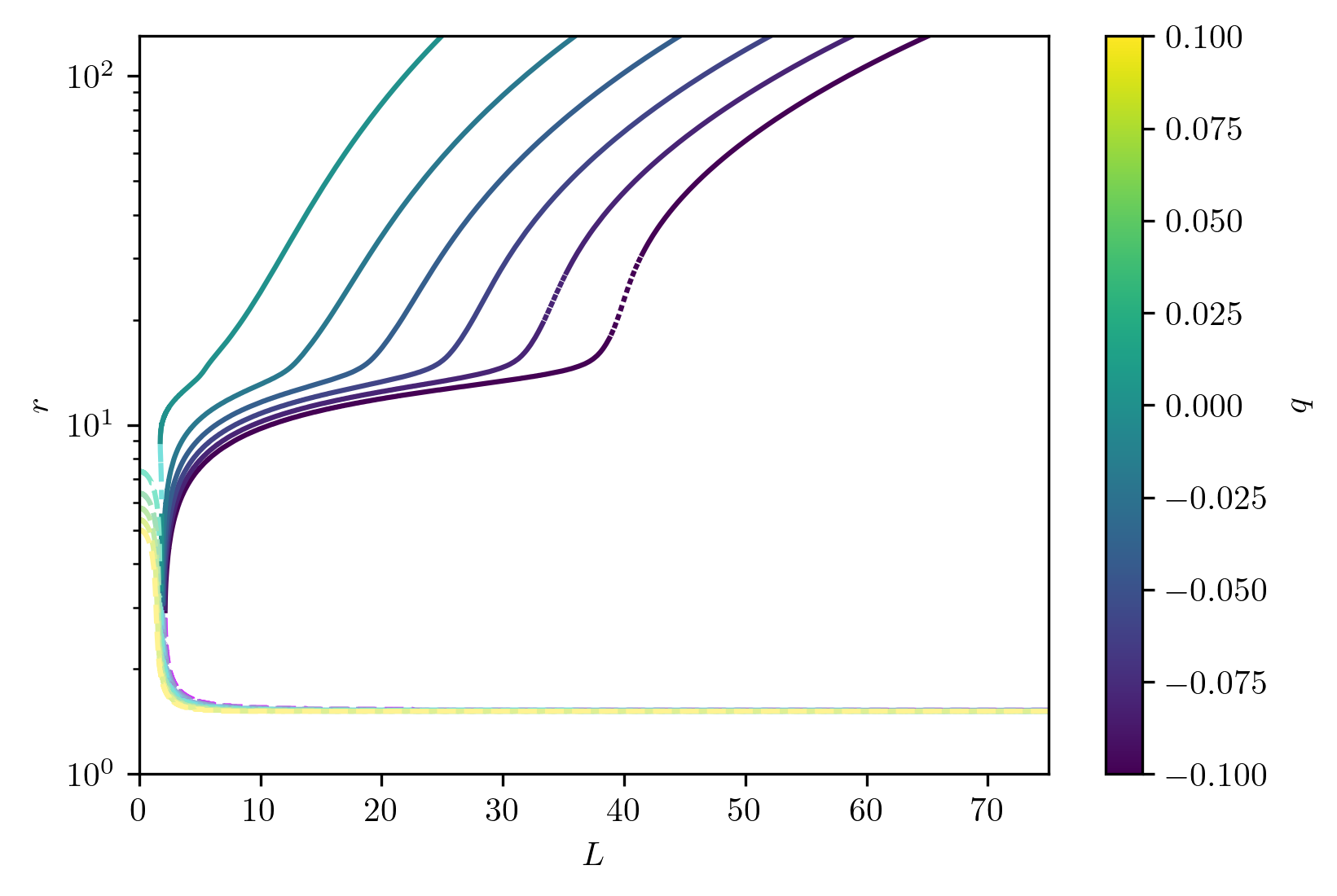}
    \includegraphics[width=0.4\linewidth]{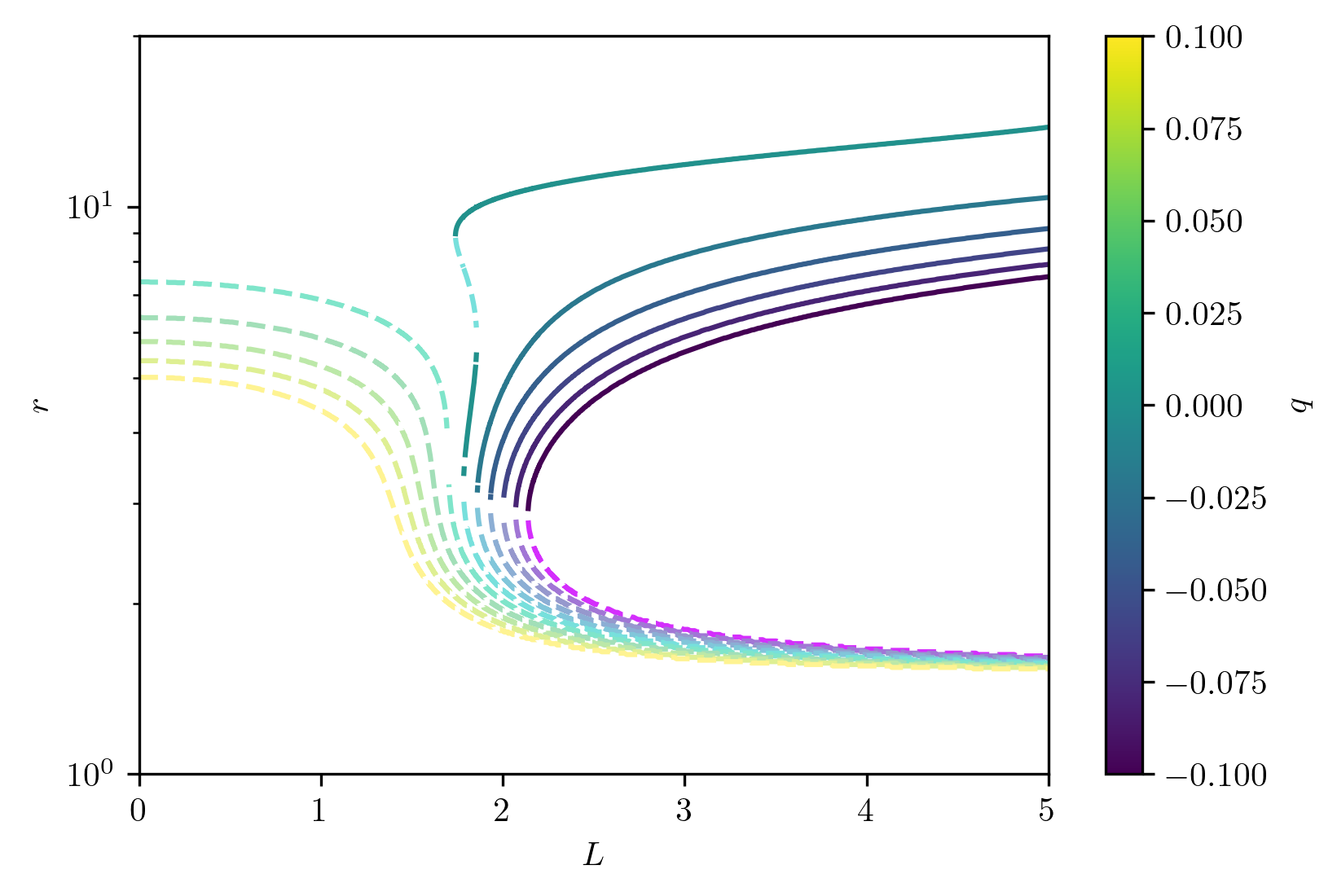}
    \caption{We show the radii of stable (solid) and unstable (dashed) circular orbits of (un)charged particles in dependence of the angular momentum $L$ in the
    space-time of a dyonically charged black hole ($k=1$) with scalar hair for $\alpha=0.001$, $g=0.05$, $\phi(r_h)=\phi_h=1.4$. The figure on the right shows the details for $L\in [0:5]$. Note that the gaps in the curves exist due to the fact that numerical accuracy does sometimes not allow to determine the exact value of the radius of the circular orbit.}
    \label{fig:circular_charged_g_0_05_k_1_ph_1_4}
\end{figure}

We have also studied the case $q\neq 0$. The results are shown in Fig.~\ref{fig:circular_charged_g_0_03_k_0_ph_5_6} (right). Comparison with the $k=0$ case
demonstrates that the pattern is similar, but that there are some quantitative differences.
We find that the static orbits of positively charged particles have larger radii in the dyonically
charged black hole space-time. Also, the radii of the stable circular orbits are larger in the $k=1$ case as compared to the $k=0$ case. This is also true for the ISCOs. On the other hand, the value of the angular momentum $L$ on the ISCO
shows weaker dependence on the charge of the particle $q$ for $k=1$.

When comparing the dependence of $r_{ISCO}/M$ on the mass to charge ratio $K/M$ of the solution,
see Fig.~\ref{fig:rISCO_EISCO_all} (left), we find that at $g=0.03$, while the radii of the ISCOs for $k=0$ are smaller than those for $k=1$, that $E_{\rm ISCO}$ is larger for $k=0$ as compared to $k=1$. This means that the infall towards the dyonic black hole generates more radiation.

We have then studied the case $g=0.05$. For large values of $\phi_h$ we find a behaviour similar to that in the RN case, while for smaller values of the scalar field on the horizon, we find additional pairs of circular orbits as well as static orbits with $L=0$ even in the case of uncharged test particles. Our result for uncharged and charged particles are
shown in Fig.~\ref{fig:circular_charged_g_0_05_k_1_ph_1_4}. 
We observe a much stronger dependence on the charge $q$ when considering the stable circular orbits as compared to the unstable orbits. Moreover, for negative $q$, we find that the behaviour is very similar to the $q=0$ case. The radius of the ISCO decreases slightly and has larger angular momentum. For sufficiently positive $q$, we find that stable circular orbits are no longer possible and that now static (unstable) orbits exist. The radius of this static orbit decreases when increasing $q$. An interesting phenomenon that also exists in the $g=0.03$ case, but is more apparent here is that at the transition from stable circular being possible  to being impossible, we find that orbits on the stable branch become unstable for a small interval in $L$, see the fifth curve from the left in Fig. \ref{fig:circular_charged_g_0_05_k_1_ph_1_4} (right).

Comparing the values of $r_{ISCO}/M$ in function of $K/M$, we find that for solutions that are not close to the corresponding RN limit, $r_{ISCO}/M$ is always largest for larger $g$. Correspondingly, the cconvertion into gravitational energy is less efficient when $g$ is larger.

\subsection{Particle collisions}

Again, we have studied both uncharged as well as charged particles in the space-time of charged black holes with scalar hair. We will first discuss the electrically charged black hole space-times and then compare with the dyonically charged case. The particles are assumed to be at rest at infinity, hence \(E=1\) for uncharged particles and \(E=1-qV_{\infty}\) for particles with electric charge.\\

\subsubsection{$k=0$: Electrically charged black holes with scalar hair}

\begin{figure}[h!]
\begin{center}
{\includegraphics[width=6.2cm]{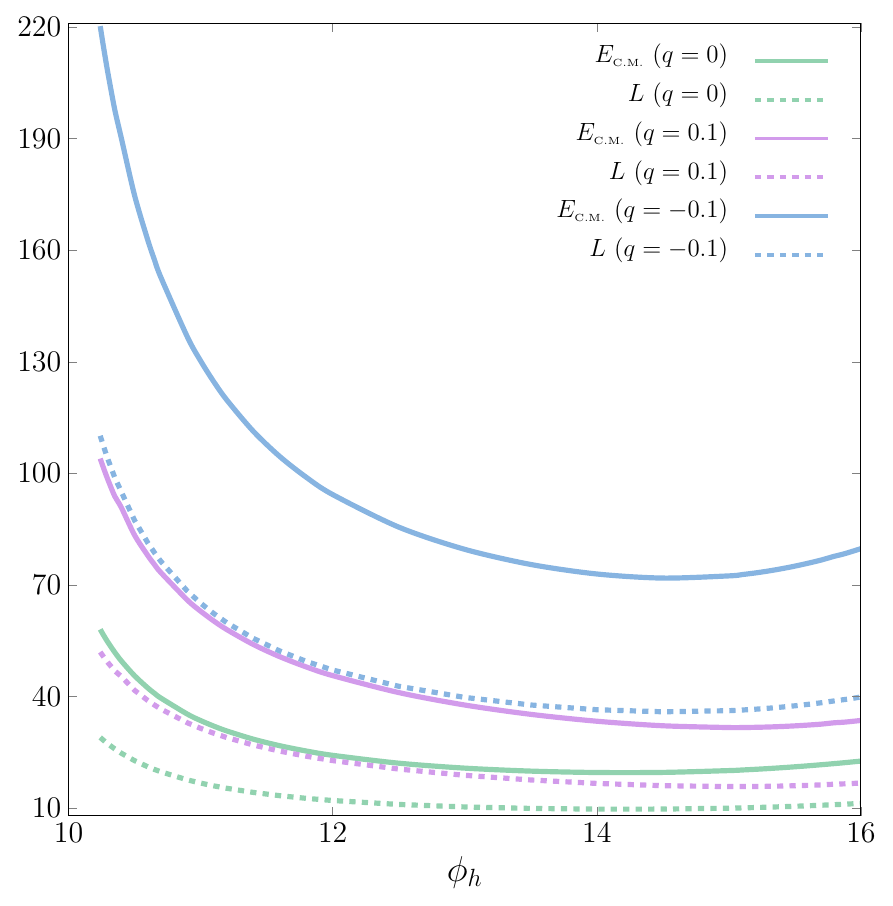}}
{\includegraphics[width=6.5cm]{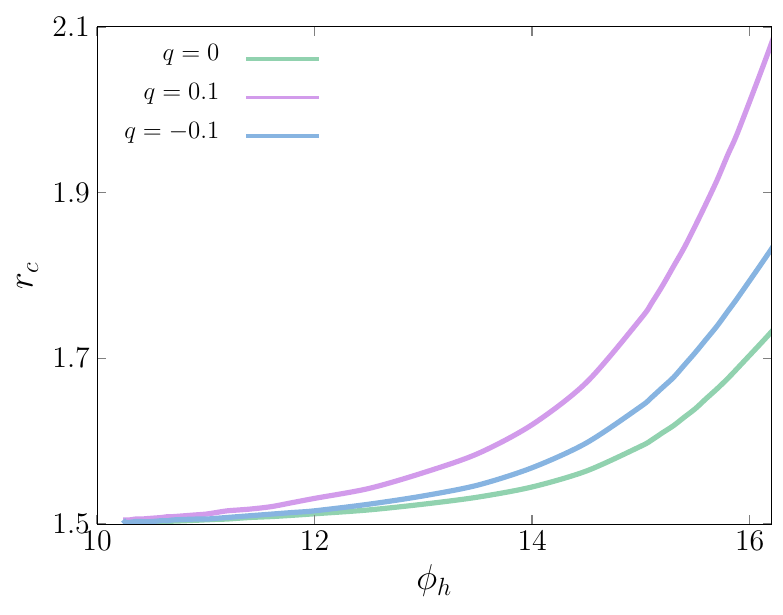}}
\vspace{-0.05cm}
\caption{{\it Left}: We show the dependence of the center-of-mass energy \(E_{\text{\tiny{C.M.}}}\) near \(r_h\) on \(\phi_h\) for uncharged particles ($q=0$), and charged particles with $q=0.1$ and $q=-0.1$, respectively, in the space-time of an electrically charged black hole ($k=0$) with scalar hair for $\alpha=0.001$, $g=0.008$. We also show the value of the angular momentum $L$ for which $\dot{r}=0$.  {\it Right}: For the same solutions we show the dependence of the critical radius $r_c$ for which $\dot{r}=0$ on the value of the scalar field at the horizon $\phi_h$ for (un)charged particles which are at rest at infinity.
\label{fig:collisions_k0_g_0_008}
}
\end{center}
\end{figure}

We have first studied the case $g=0.008$. Our results are given in Fig.~\ref{fig:collisions_k0_g_0_008}, where we show the center-of-mass energy \(E_{\text{\tiny{C.M.}}}\) and the critical angular momentum $L$ at which $\dot{r}=0$ (left) as well as the critical radius $r_c$ at which $\dot{r}=0$ (right), in dependence of the value of the scalar field on the horizon $\phi(r_h)=\phi_h$. 
We observe that collisions near the horizon are no longer possible when $\phi_h > 16$ as $r_c$ is considerably larger than $r_h=1$, hence we do not compute the center-of-mass energy for $\phi_h > 16$. For each scalar field configuration, the center-of-mass energy has then been computed by considering two particles with identical charge but opposite signs of the critical angular momentum.

For all three charges $q$, the center-of-mass energy \(E_{\text{\tiny{C.M.}}}\) and the critical angular momentum $L$ decrease as the scalar field on the horizon increases. The rate of decrease is initially rapid for smaller
values of $\phi_h$, indicating that the influence of the scalar field becomes weaker at larger values of $\phi_h$. Near the upper end of the parameter range, both quantities exhibit a slight increase.

\begin{figure}[h!]
\begin{center}
{\includegraphics[width=6.2cm]{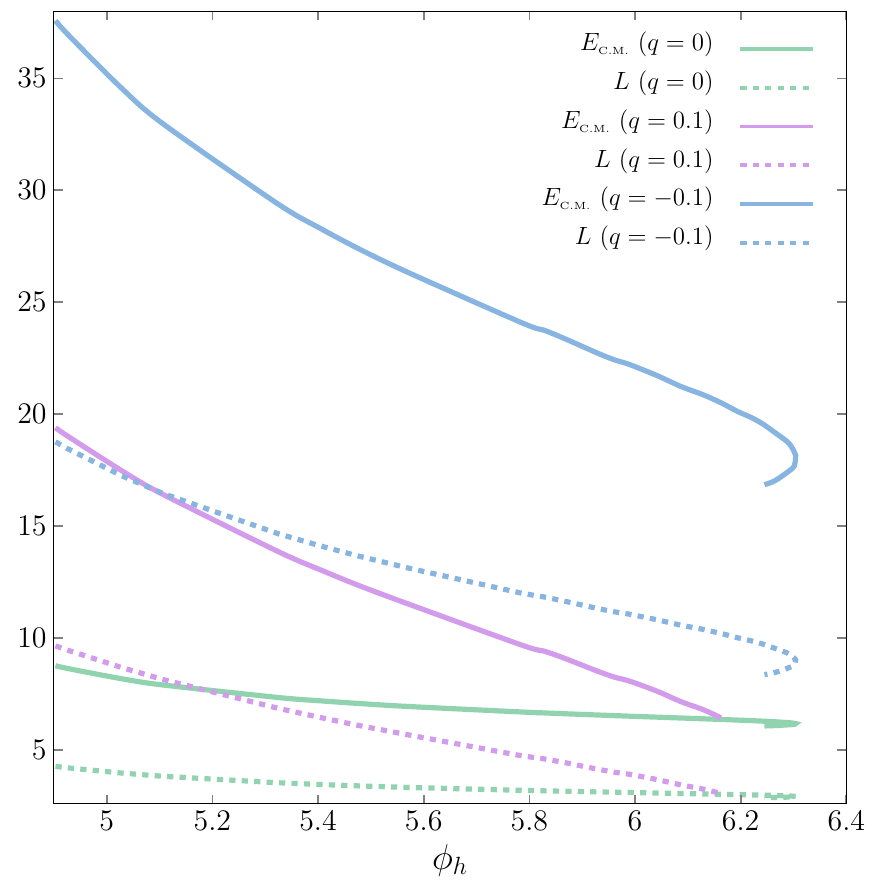}}
{\includegraphics[width=6.5cm]{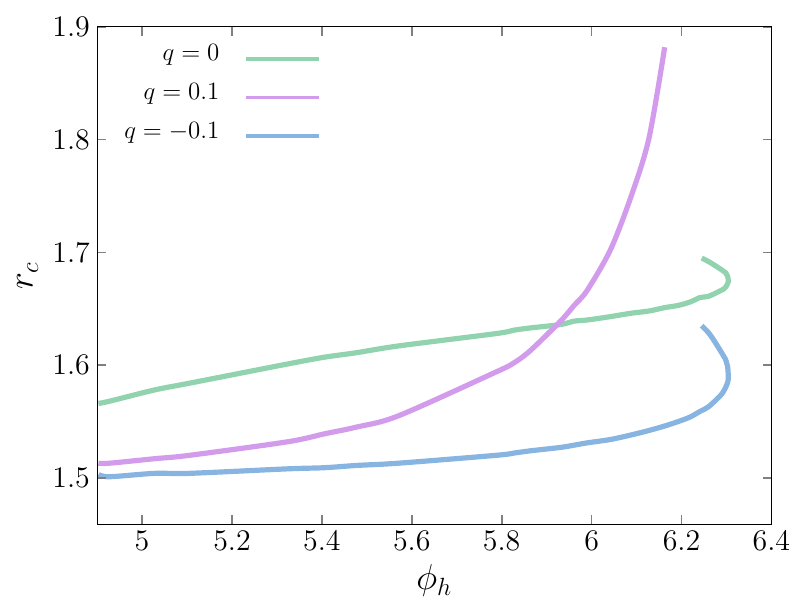}}
\vspace{-0.05cm}
\caption{{\it Left}: We show the dependence of the center-of-mass energy \(E_{\text{\tiny{C.M.}}}\) near \(r_h\) on \(\phi_h\) for uncharged particles ($q=0$), and charged particles with $q=0.1$ and $q=-0.1$, respectively, in the space-time of an electrically charged black hole ($k=0$) with scalar hair for $\alpha=0.001$, $g=0.03$. We also show the value of the angular momentum for which $\dot{r}=0$. {\it Right}: For the same solutions we show the dependence of the critical radius $r_c$ for which $\dot{r}=0$ on the value of the scalar field at the horizon $\phi_h$ for (un)charged particles which are at rest at infinity.
\label{fig:collisions_k0_g_0_03}
}
\end{center}
\end{figure}

We have also studied the case $g=0.03$. The results are shown in Fig.~\ref{fig:collisions_k0_g_0_03}.

The numerical results demonstrate that increasing the scalar field at the horizon leads to
a decrease in both the center-of-mass energy and the critical angular momentum,
indicating that a larger scalar field on the horizon suppresses the conditions required for high energy particle collisions. An additional feature is observed for $q=0$ and $q=-0.1$, respectively,
where both the center-of-mass energy and the angular momentum exhibit a slight turning point
near $\phi_h\approx 6.25$. The critical radius $r_c$ generally increases with increasing scalar field value on the horizon,
indicating that the location satisfying the critical condition moves further away from the horizon
as the scalar hair becomes larger. For $q=-0.1$, the critical radius remains relatively small,
indicating that critical trajectories stay closer to the horizon. For $q=0.1$, the critical radius
grows rapidly at larger $\phi_h$, suggesting that electromagnetic repulsion becomes increasingly
important and pushes the critical orbit further from the horizon. The $q=0$ case is intermediate to the two charged cases discussed here. This indicates that the scalar hair modifies the near
horizon geometry by relocating the critical orbit away from the horizon, thereby reducing the
efficiency of particle acceleration.

Comparing the two cases, our numerical results indicate that decreasing the coupling parameter $g$ in the purely electric setting significantly enhances the Ba\~nados–Silk–West process. Smaller values of $g$ allow particles to achieve higher center-of-mass energies, particularly for negatively charged particles. Although increasing the horizon scalar field generally suppresses both the collision energy and the critical angular momentum, this suppression is not monotonic over the entire parameter space. The appearance of shallow minima followed by a slight increase suggests that the scalar field and the coupling parameter compete in determining the near-horizon structure, leading to a nontrivial dependence of the collision dynamics on $\phi_h$.

\begin{figure}[h!]
\begin{center}
{\includegraphics[width=8cm]{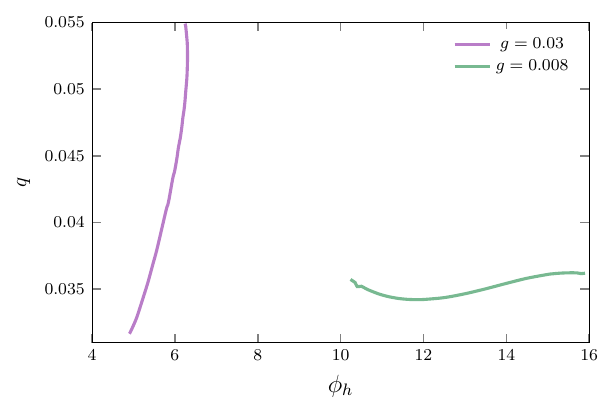}} 
{\includegraphics[width=8cm]{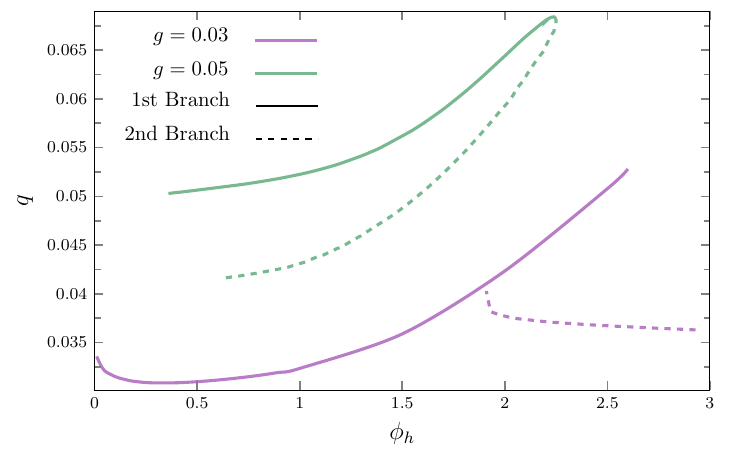} }
\vspace{-0.05cm}
\caption{{\it Left}: We show the value of the particle's charge $q=1/V_{\infty}$ in dependence of the value of the scalar field on the horizon $\phi_h$ that leads to an infinite center-of-mass energy for electrically charged black holes with scalar hair. We give the dependence for two values of g, $g=0.008$ (green) and $g=0.03$ (purple). {\it Right}: Same as left, but for dyonically charged black holes and $g=0.03$ (purple) and $g=0.05$ (green). For all solutions, we have chosen $\alpha=0.001$ and $r_h=1.0$.}
\label{fig:hairyBHcoll3}
\end{center}
\end{figure}

We do not find the center-of-mass energy to diverge in the cases shown, i.e. for uncharged particles and charged particles with $q=\pm 0.1$. However, the center-of-mass energy
can become infinite for charged particles when the energy $\tilde{E}_i=E_i+q_i V(r)$ vanishes (see (\ref{eq:com_energy})). With our choice of gauge this implies that $\tilde{E}=0$ for $q=1/V_{\infty}$. Since $V_{\infty}$ is only numerically given, such is $q$. Our results for electrically charged black holes are shown in Fig. \ref{fig:hairyBHcoll3} (left).
We find that for all solutions we need $q > 0$ in order to find infinite center-of-mass energy. Again, the qualitative behaviour is very different for the two values of $g$. For $g=0.008$, the charge $q$ that leads to infinite center-of-mass energy
first decreases when increasing $\phi_h$ until $\phi_h\approx 12$, then increases until $\phi_h\approx 15.5$ and then decreases again. For $g=0.03$, on the other hand, the charge $q$
is monotonically increasing with $\phi_h$.

\subsubsection{$k=1$: Dyonically charged black holes with scalar hair}
In the following, we have only discussed uncharged and electrically charged particles and their collision. One could also discuss dyonically charged test particles, i.e. particles
that carry both electric and magnetic charge. We have not investigated this case in detail using our numerical data, but give more details in Appendix \ref{Appendixa}.
In particular, we find that the center-of-mass energy can also diverge in this latter case and we give the condition under which it would.

Similar to the $k=0$ case, we consider particle collision in both uncharged as well as charged particles. The particles are assumed to be at rest at infinity, hence \(E=1\) for uncharged particles and \(E=1-qV_{\infty}\) for particles with electric charge.

\begin{figure}[h!]
\begin{center}
{\includegraphics[width=6.2cm]{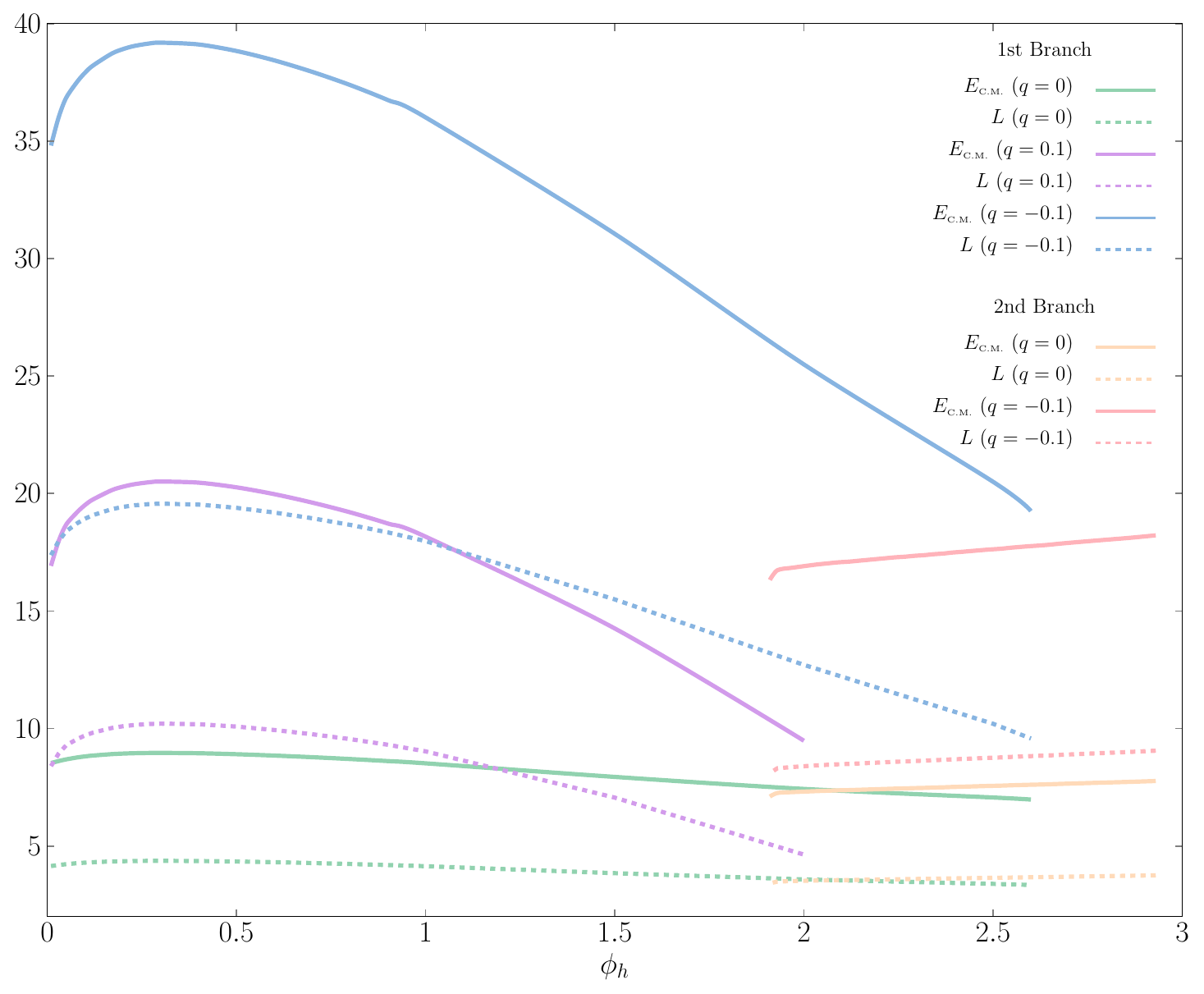}}
{\includegraphics[width=6.5cm]{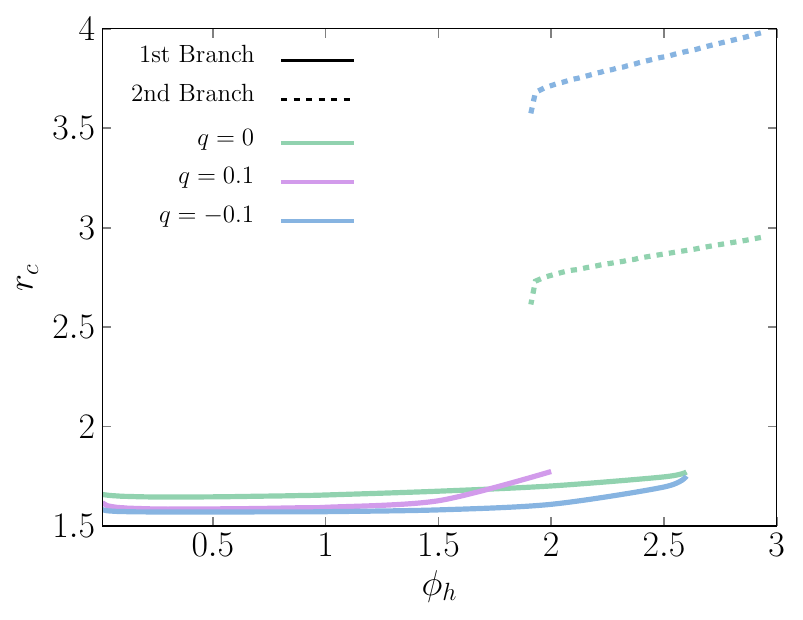}}
\vspace{-0.05cm}
\caption{{\it Left}: We show the dependence of the center-of-mass energy \(E_{\text{\tiny{C.M.}}}\) near \(r_h\) on \(\phi_h\) for uncharged particles ($q=0$), and charged particles with $q=0.1$ and $q=-0.1$, respectively, in the space-time of a dyonically charged black hole ($k=1$) with scalar hair for $\alpha=0.001$, $g=0.03$. We also give the value of the angular momentum $L$ for which $\dot{r}=0$. {\it Right}: For the same solutions we show the dependence of the critical radius $r_c$ for which $\dot{r}=0$ on the value of the scalar field at the horizon $\phi_h$ for (un)charged particles which are at rest at infinity.
\label{fig:collisions_k1_g_0_03}
}
\end{center}
\end{figure}

The values of $L$ at $\dot{r}=0$ near the horizon as well as the center-of-mass energy $E_{\text{\tiny{C.M.}}}$ in dependence of the scalar field on the horizon, $\phi(r_h)=\phi_h$ are given in Fig.~\ref{fig:collisions_k1_g_0_03} for $g=0.03$. The numerical solutions separate into two distinct branches. Along the first branch, both the
center-of-mass energy and the critical angular momentum become quite dependent on the scalar field.
For all charge configurations, the center-of-mass energy initially increases slightly for small
values of $\phi_h$, reaches a maximum, and subsequently decreases as the scalar field continues to
grow. A similar trend is observed for the critical angular momentum, although its variation is more
gradual. On the second branch both the center-of-mass energy and the critical angular momentum
increase gradually with increasing $\phi_h$. On the first branch, the scalar field
initially enhances the collision energy before becoming the dominant suppressing factor, leading to
a steady decrease in both the center-of-mass energy and the critical angular momentum. The second branch, however, exhibits the opposite
trend. Here, increasing the scalar field leads to a gradual enhancement of both the center-of-mass
energy and the critical angular momentum, implying that the near-horizon geometry evolves
differently along this family of solutions.
Comparing this to the $k=0$, $g=0.03$ case, see above, we obviously find that the presence of the magnetic charge influences the collisions of particles. The center-of-mass energy can become larger when the additional
magnetic charge is present - for smaller values of $\phi_h$ on the horizon.

\begin{figure}[h!]
\begin{center}
{\includegraphics[width=6.2cm]{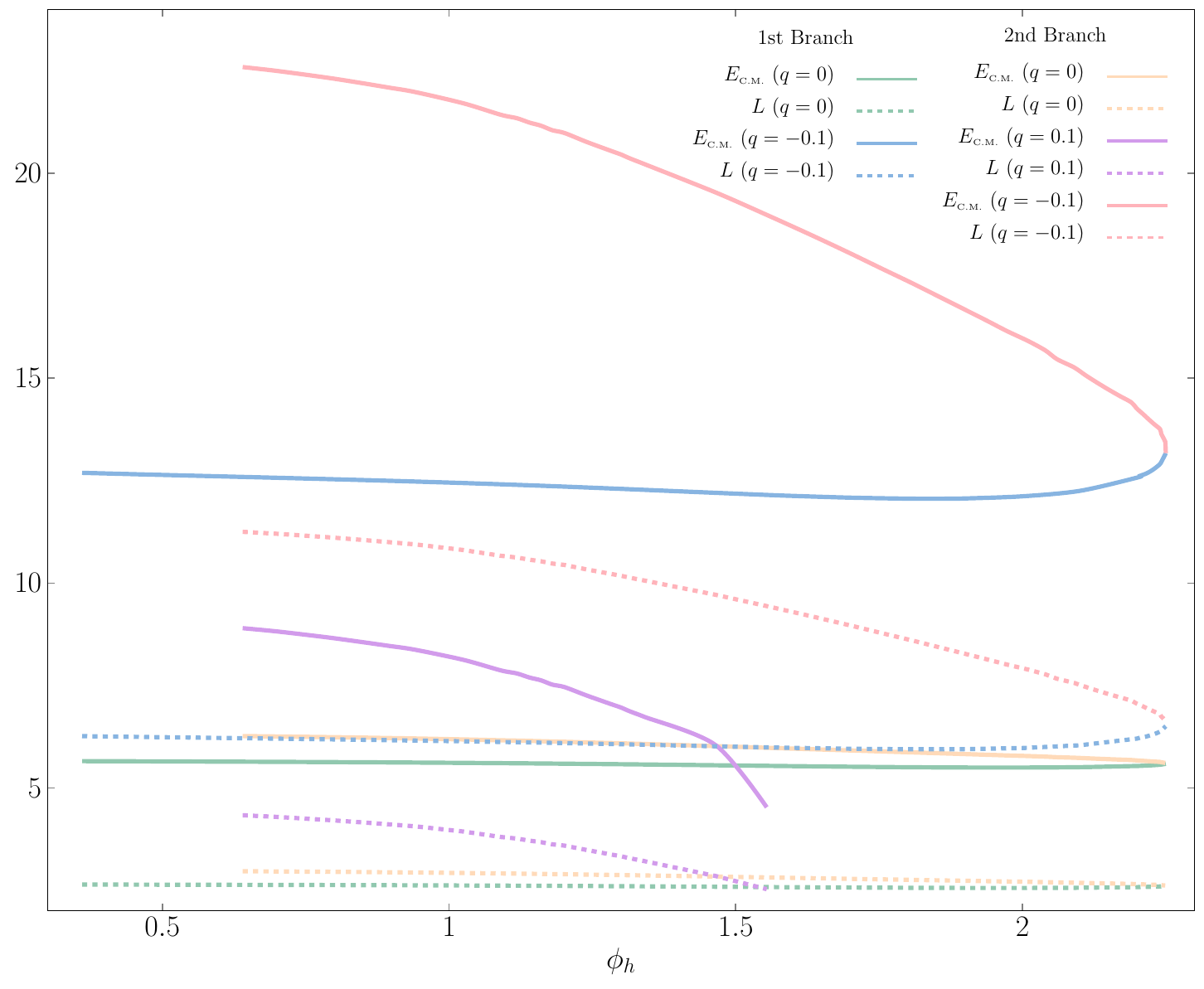}}
{\includegraphics[width=6.5cm]{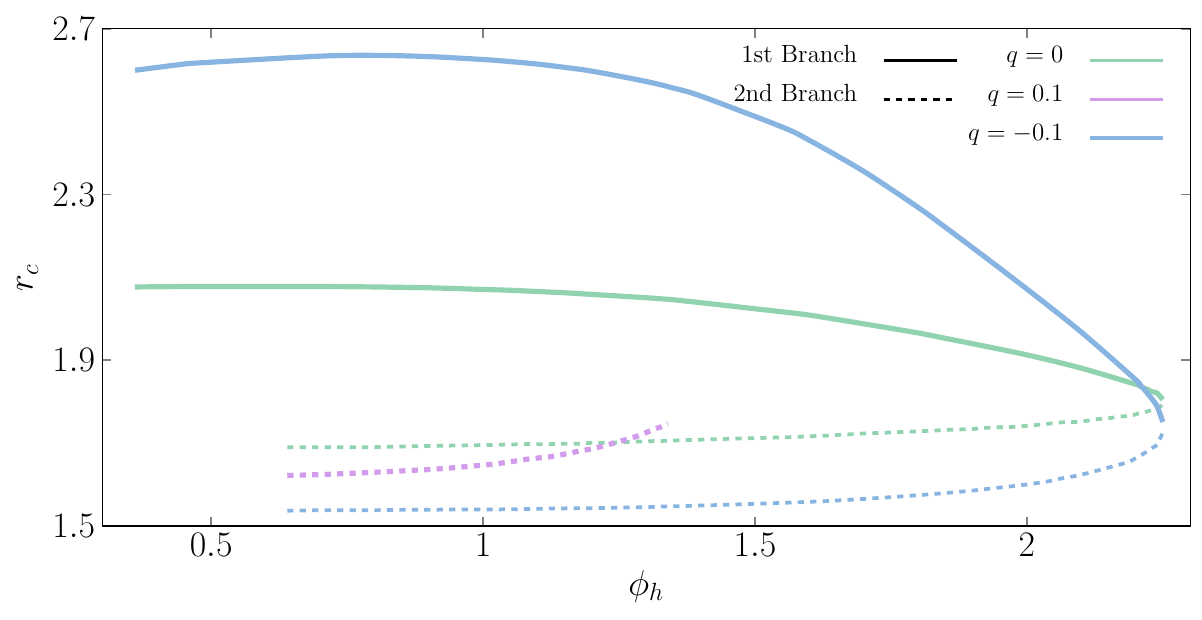}}
\vspace{-0.05cm}
\caption{{\it Left}: We show the dependence of the center-of-mass energy \(E_{\text{\tiny{C.M.}}}\) near \(r_h\) on \(\phi_h\) for uncharged particles ($q=0$), and charged particles with $q=0.1$ and $q=-0.1$, respectively, in the space-time of a dyonically charged black hole ($k=1$) with scalar hair for $\alpha=0.001$, $g=0.05$. We also give the value of the angular momentum $L$ for which $\dot{r}=0$.  {\it Right}: For the same solutions we show the dependence of the critical radius $r_c$ for which $\dot{r}=0$ on the value of the scalar field at the horizon $\phi_h$ for (un)charged particles which are at rest at infinity.
\label{fig:collisions_k1_g_0_05}
}
\end{center}
\end{figure}

To understand how $g$ influences these results, we have also studied the case $g=0.05$. The results are shown in Fig.~\ref{fig:collisions_k1_g_0_05}.
Along the first branch, both the center-of-mass energy and the critical angular momentum slowly
decrease with increasing $\phi_h$. The growth is smooth for all particle charges, indicating that the scalar field enhances the particle acceleration process within this family of solutions. On the
second branch, as the scalar field increases, both the center-of-mass energy and the critical
angular momentum decrease monotonically for all charge configurations. The numerical analysis shows
that collisions between two positively charged particles occur only on the second branch and are
confined to a limited range of $\phi_h$. Along the first branch, the critical radius decreases with
increasing $\phi_h$ for all particle charges, indicating that stronger scalar hair shifts the
critical orbit towards the horizon. The second branch displays the opposite behavior. Here, the
critical radius increases as the scalar field on the horizon increases, implying that the critical
orbit moves further away from the horizon.

Analogous to the purely electric case, infinite center-of-mass energy can be obtained for dyonically charged black holes by choosing \(q=1/V_{\infty}\). The corresponding numerical results are displayed in Fig. \ref{fig:hairyBHcoll3} (right). On the first branch (solid curves), the critical charge generally increases with increasing \(\phi_h\). In the \(g=0.03\) case, \(q\) initially decreases slightly, reaches a minimum near \(\phi_h\approx0.03\), and then grows monotonically as the scalar field increases. In contrast, for \(g=0.05\), the required charge is systematically larger and rises more rapidly with \(\phi_h\).

\section{Conclusions}
In this paper, we have studied how massive particle motion changes when scalar fields are present in the space-time of charged, spherically symmetric black holes. We have
discussed four families of solutions of black holes, two electrically charged and two dyonically charged.

Circular motion is possible in this case and while for Reissner-Nordstr\"om only a single pair of circular orbits exists, when scalar hair is considered, we find that often two pairs of circular orbits exist, two unstable and two stable circular orbits. For each pair, the unstable and stable branch meets at an ISCO at a critical value of the angular momentum. Depending on the value of the scalar field on the horizon, we find that static orbits with $L = 0$ are possible. This is unique to the solutions with non-vanishing scalar fields and is not possible for the RN space-time.
When test particles are given charge, we find that for negative particles (of opposite sign to the black hole), similar behaviour occurs to when the value of the scalar field on the horizon is changed. For sufficiently positive charge, however, stable orbits are lost and only a single unstable orbit occurs for each $L$. 

We have also investigated particle collisions in both the Reissner–Nordstr\"om space-time and the space-time of charged black holes with scalar hair to determine the conditions under which arbitrarily large center-of-mass energies can be achieved near the event horizon. For the Reissner–Nordstr\"om spacetime, we have shown that an infinite center-of-mass energy is possible only when one particle has zero angular momentum and the other possesses the corresponding critical angular momentum. This condition can be realised when both particles are charged or when one particle is charged and the other is neutral.

For charged black holes with scalar hair, we analised collisions between particles carrying identical charges (neutral, positive, or negative) but opposite signs of angular momentum. This allowed us to determine both the collision radius and the critical angular momentum associated with each configuration. Our results demonstrate that large center-of-mass energies can be achieved for both electrically charged and dyonic black holes. However, infinite center-of-mass energies arise only when one of the colliding particles has a charge \(q=1/V_{\infty}\) while the second particle possesses an arbitrary charge or remains neutral. As in the Reissner–Nordstr\"om case, this requires one particle to have zero angular momentum and the other to carry the critical angular momentum. Our results show that the presence of scalar hair does not remove the need for finely tuned particle parameters to achieve the Ba\~nados–Silk–West effect. Instead, the scalar field modifies the conditions under which critical particle trajectories exist, while preserving the underlying requirement that one of the colliding particles must satisfy a critical set of conserved quantities. This highlights the important role played by both the particle charge and angular momentum in determining the efficiency of particle acceleration in the vicinity of charged black holes.
\\
\\
{\bf Acknowledgements} Katherine Horton is supported by the Engineering and Physical Sciences Research Council under grant number EP/W524335/1.
We thank Jutta Kunz for pointing out reference \cite{Wei:2023bgp} to us.


\clearpage

\appendix

\section{No static orbits in the space-time of a Reissner-Nordstr\"om black hole}
\label{Appendixb}
In the following, we will demonstrate that no static spheres exist in the Reissner-Nordstr\"om (RN) space-time and that, hence, the presence of the scalar field is crucial to allow for static spheres to exist.

Next to the condition $\dot{r}=0$ for a circular orbit, we require $\dot{\theta}=0$ as well as $\dot{\varphi}=0$ for a static orbit. The former implies $\theta={\rm constant}$, the latter $L=0$. The effective potential (\ref{eq:rdot}) then reads $V_{\rm eff}=-(E+qV)^2 + N$. Requiring
$V_{\rm eff}=V'_{\rm eff}=0$ we find
\begin{equation}
4N = \left(\frac{N'}{q V'}\right)^2  
\end{equation}
and inserting the expressions for the RN solution, we find the following quadratic equation in $r$~:
\begin{equation}
\label{eq:static_RN}
r^2\left(q^2 Q^2 - M^2\right) - 2 M r(q^2 Q^2 - K^2) + q^2 K^2 Q^2 - K^4 = 0 
\end{equation}
where $K^2=\alpha(Q^2 +Q_m^2)$. We can now set $M\equiv 1$ without loss of generality by rescaling $r\rightarrow Mr$, $Q\rightarrow MQ$ and $K\rightarrow MK$. Letting $K² = \alpha (Q^2 + Q_m^2)$ and requiring that (\ref{eq:static_RN}) has at least one real zero, we find that
\begin{equation}
\left[q^2 Q^2 - \alpha(Q^2+Q_m^2) \right]^2 - (q^2 Q^2 -1)\left[\alpha q^2 Q^2 (Q^2 +Q_m^2) -\alpha^2 (Q^2+Q_m^2)^2\right]  \geq 0 \ .
\end{equation}
Dividing this inequality by $q^2 Q^2$ and re-arranging gives
\begin{equation}
\alpha\left(Q^2 + Q_m^2\right) \geq 1 
\end{equation}
which (with $M=1$) can only be fulfilled for the extremal RN black hole solution, i.e. for $M=\alpha(Q^2+ Q_m^2)$. It is then easy to see that (\ref{eq:static_RN}) becomes $r^2 - 2Mr + M^2=0$ such that the static sphere corresponds to the extremal horizon with radius $r_h=M=\alpha\left(Q^2 + Q_m^2\right)$.

\section{Collisions of electrically and magnetically charged particles}
\label{Appendixa}
Motion of particles which hold both electric and magnetic charges in the space-time of a dyonically charged RN solution has been studied in \cite{Grunau:2011gd}. Here, we extend the discussion to the collision of particles and also include the possibility of a scalar field.
In the spacetime (\ref{eq:metric}) we find~:
\begin{equation}
    \tilde{F}_{tr}=i\frac{Q_m\sigma(r)}{r^2}  \ \ , \ \  \tilde{F}_{\theta\varphi}=-i\frac{r^2V'(r)\sin\theta }{\sigma(r)}  \ ,
    \end{equation}
where the dual field strength is given by \(\tilde{F}^{\mu\nu}=\frac{i}{2\sqrt{-g}}\varepsilon^{\mu\nu\rho\sigma}F_{\rho\sigma}\). In addition to the gauge potential in (\ref{electromagnetic}), the non-vanishing components of the gauge potential \(\tilde{A}_{\mu}\) are~:
\begin{equation}
\tilde{A}_{t}=-iQ_m\int^r\frac{\sigma(\tilde{r})}{\tilde{r}^2}d\tilde{r}+C_t \ \ , \ \  \tilde{A}_{\varphi}=iQ\cos\theta \ ,
\end{equation}
where \(C_t\) is gauge constant and \(Q=r^2V'(r)/\sigma(r)\). The Hamilton–Jacobi equation governing a particle endowed with electric charge \(q\) and magnetic charge \(q'\) is
\begin{equation}
\frac{\partial S}{\partial\tau}=\frac{1}{2}g^{\mu\nu}\left(\frac{\partial S}{\partial x^{\mu}}-qA_{\mu}+iq'\tilde{A}_{\mu}\right)\left(\frac{\partial S}{\partial x^{\nu}}-qA_{\nu}+iq'\tilde{A}_{\nu}\right).\end{equation}
Restricting to the equatorial plane, the 4-velocities of the two particles are~:
\begin{equation}
u^{\mu}_{(i)}=\Bigg(\frac{\tilde{E}_i - q_i'Q_m \displaystyle\int^r 
\frac{\sigma(\tilde{r})\, d\tilde{r}}{\tilde{r}^2}-iq_i'C_t}{N\sigma^2},\sqrt{
\frac{\left(\tilde{E}_i - q_i'Q_m \displaystyle\int^r 
\frac{\sigma(\tilde{r})\, d\tilde{r}}{\tilde{r}^2}-iq_i'C_t\right)^2}{\sigma^2}
- N \left(1 + \frac{L_i^2}{r^2}\right)},0,\frac{L_i}{r^2}\Bigg) \ ,
\end{equation}
where \(\tilde{E}_i=E_i+q_iV(r)\) and using the normalization $\dot{x}_{\mu}\dot{x}^{\mu}=-1$ we find~:
\begin{equation}
E_i= \sigma\sqrt{ \dot{r}^2 + N  r^2\dot{\theta}^2 + \frac{N(L    -q_i Q_m \cos\theta-q_i'Q\cos\theta)^2}{r^2\sin^2\theta} + N} - q_i V+q_i'Q_m\int^{\infty}_r\frac{\sigma(\tilde{r})d\tilde{r}}{\tilde{r}^2}+iq'_iC_t \ .
\end{equation}
Choosing the gauge to be at infinity, the energy for a particle at rest at infinity is \(E_i=1-q_iV_{\infty}\) for \(i=1, 2\). The center-of-mass energy near the horizon becomes~:
\begin{eqnarray}
\frac{E^2_{\text{\tiny{C.M.}}}}{2m_0^2}
&=& 1+\frac{1}{2}\bigg(\tilde{E}_1-q_1'Q_m\int^{\infty}_{r_h}\frac{\sigma(\tilde{r})d\tilde{r}}{\tilde{r}^2}\bigg)\bigg(\tilde{E}_2-q_2'Q_m\int^{\infty}_{r_h}\frac{\sigma(\tilde{r})d\tilde{r}}{\tilde{r}^2}\bigg)^{-1}\bigg(1+\frac{L^2_2}{r_h^2}\bigg) \nonumber \\
&+&\frac{1}{2}\bigg(\tilde{E}_1-q_1'Q_m\int^{\infty}_{r_h}\frac{\sigma(\tilde{r})d\tilde{r}}{\tilde{r}^2}\bigg)^{-1}\bigg(\tilde{E}_2-q_2'Q_m\int^{\infty}_{r_h}\frac{\sigma(\tilde{r})d\tilde{r}}{\tilde{r}^2}\bigg)\bigg(1+\frac{L^2_1}{r_h^2}\bigg)-\frac{L_1L_2}{r_h^2} \ .  
\end{eqnarray}
The above expression suggests that it might be possible to obtain an infinite \(E_{\text{\tiny{C.M.}}}\) provided that \(\int^{\infty}_{r_h}\frac{\sigma(\tilde{r})d\tilde{r}}{\tilde{r}^2}\) is non-zero on the horizon and that \(\tilde{E}_i=q'_iQ_m\int^{\infty}_{r_h}\frac{\sigma(\tilde{r})d\tilde{r}}{\tilde{r}^2}\) for either \(i=1\) or \(i=2\).


\end{document}